\def\m{{\rm\,m}}
\def\km{{\rm\,km}}

\def\yr{{\rm\,yr}}
\def\Myr{{\rm\,Myr}}

\documentclass[preprint2]{proto}
\usepackage{times}
\usepackage{textcomp}
\usepackage{wasysym}
\usepackage{graphicx}
\usepackage{epstopdf}
\newcommand{\refs}{\par\noindent\hangindent=1pc\hangafter=1}
\begin{document}

\title{\textbf{\LARGE Binaries in the Kuiper Belt}}

\author {\textbf{\large Keith S.~Noll}}
\affil{\small\em Space Telescope Science Institute}

\author {\textbf{\large William M.~Grundy}}
\affil{\small\em Lowell Observatory}

\author {\textbf{\large Eugene I.~Chiang}}
\affil{\small\em University of California, Berkeley}

\author {\textbf{\large Jean-Luc Margot}}
\affil{\small\em Cornell University}

\author {\textbf{\large Susan D.~Kern}}
\affil{\small\em Space Telescope Science Institute}

\vspace{1 cm}

\author {{Accepted for publication in {\it The Kuiper Belt} }}
\affil{{\it University of Arizona Press, Space Science Series}}

\begin{abstract}
\baselineskip = 11pt
\leftskip = 0.65in 
\rightskip = 0.65in
\parindent=1pc
{\small Binaries have played a crucial role many times in the history of modern astronomy and are doing so again in the rapidly evolving exploration of the Kuiper Belt.  The large fraction of transneptunian objects that are binary or multiple,  48 such systems are now known, has been an unanticipated windfall.  Separations and relative magnitudes measured in discovery images give important information on the statistical properties of the binary population that can be related to competing models of binary formation.  Orbits, derived for 13 systems, provide a determination of the system mass.  Masses can be used to derive densities and albedos when an independent size measurement is available.  Angular momenta and relative sizes of the majority of binaries are consistent with formation by dynamical capture.  The small satellites of the largest transneptunian objects, in contrast, are more likely formed from collisions.  Correlations of the fraction of binaries with different dynamical populations or with other physical variables have the potential to constrain models of the origin and evolution of the transneptunian population as a whole.  Other means of studying binaries have only begun to be exploited, including lightcurve, color, and spectral data.  Because of the several channels for obtaining unique physical information, it is already clear that binaries will emerge as one of the most useful tools for unraveling the many complexities of transneptunian space.
 \\~\\~\\~}
 
\end{abstract}

\section{\textbf{HISTORY AND DISCOVERY}}

Ever since Herschel noticed, two hundred years ago, that gravitationally bound stellar binaries exist, the search for binaries has followed close on the heels of the discovery of each new class of astronomical object.   The reasons for such searches are, of course, eminently practical.  Binary orbits provide determinations of system mass, a fundamental physical quantity that is otherwise difficult or impossible to obtain.  The utilization of binaries in stellar astronomy has enabled countless applications including Eddington's landmark mass-luminosity relation.  Likewise, in planetary science, bound systems have been extensively exploited; they have been used, for example, to determine the masses of planets and to make Roemer's first determination of the speed of light.  The statistics of binaries in astronomical populations can be related to formation and subsequent evolutionary and environmental conditions.  

Searches for bound systems among the small body populations in the solar system has a long and mostly fruitless history that has been summarized in several recent reviews ({\it Merline et al.}, 2002; {\it Noll}, 2004, 2006; {\it Richardson and Walsh}, 2005).  But the discovery of the {\em second} transneptunian binary (TNB), 1998 WW$_{31}$ ({\it Veillet et al.}, 2001) marked the start of a landslide of discovery that shows no signs of abating.

\begin{figure*}
 \epsscale{1.}
 \plotone{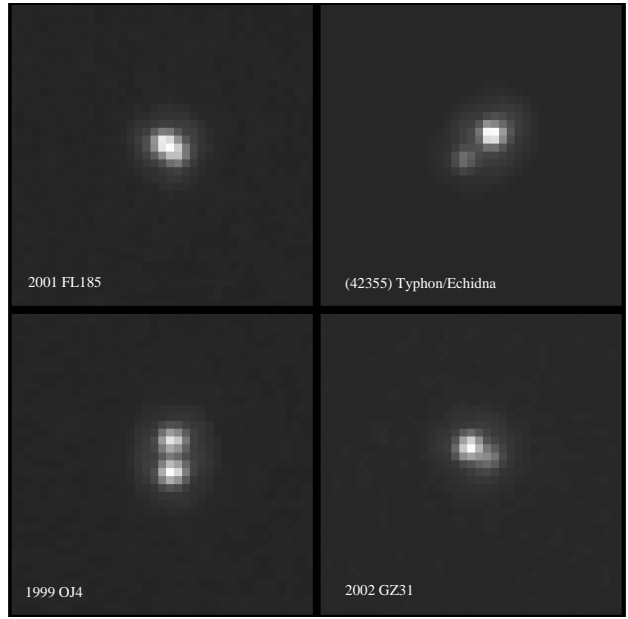}
 \caption{\small Images of 2001 FL$_{185}$ (top-left), (42355) Typhon/Echidna (top-right), 1999 OJ$_4$ (bottom-left), and 2002 GZ$_{31}$ (bottom-right).  The images shown are each combinations of four separate 300 s exposures taken with the High Resolution Camera on HST.  The dithered exposures have been combined using multidrizzle.  The images are shown with a linear grayscale normalized to the peak pixel.  The pixels in the drizzled images are 25 milliarcsec on a side in a non-distorted coordinate frame.  Each of the four panels is 1 arcsec square.  Images are oriented in detector coordinates.}  
 \end{figure*}

\bigskip
\noindent
\textbf{1.1 Discovery and Characterization of Charon}
\bigskip

The first example of what we would now call a TNB was discovered during a very different technological epoch than the present, prior to the widespread astronomical use of CCD arrays. The discovery of Charon ({\it Christy and Harrington}, 1978) on photographic plates taken for Pluto astrometry heralded a spectacular flourishing of Pluto science, and offered a glimpse of the tremendous potential of TNBs to contribute to Kuiper belt science in general.

Charon's orbit ({\it Christy and Harrington}, 1978, 1980) revealed the system mass, which up until then had been estimated via other methods, with wildly divergent results.  About the same time, spectroscopy revealed the presence of methane on Pluto ({\it Cruikshank et al}, 1976) indicating a high albedo, small size, and the possible existence of an atmosphere.  Observations of occultations of stars by Pluto confirmed its small size ({\it  Millis et al.}, 1993) and enabled direct detection of the atmosphere ({\it Hubbard et al.}, 1988).

The orbit plane of Charon, as viewed from the Earth, was oriented edge-on within a few years of Charon's discovery.  This geometry happens only twice during Pluto's 248 year orbit, so its occurrence just after Charon's discovery was fortuitous.  Mutual events, when Charon (or its shadow) passed across the face of Pluto, or Pluto (or its shadow) masked the view of Charon, were observable from 1985 through 1990 ({\it Binzel and Hubbard}, 1997).  From the timing of these events and from the changes in observable flux during them, much tighter constraints on the sizes and albedos of Pluto and Charon were derived ({\it e.g. Young and Binzel}, 1994).  Mutual events also made it possible to distinguish the surface compositions of Pluto and Charon ({\it e.g. Buie et al.}, 1987; {\it Fink and DiSanti}, 1988), by comparing reflectance spectra of the two objects blended together with spectra of Pluto alone, when Charon was completely hidden from view.  From subtle variations in flux as Charon blocked different regions of Pluto's surface (and vice versa), maps of albedo patterns on the faces of the two objects were constructed ({\it e.g. Buie et al.}, 1997; {\it Young et al.}, 1999).  The mutual events are now over and will not be repeated during our lifetimes, but telescope and detector technology continue their advance.  For relatively well-separated, bright TNBs like Pluto and Charon, it is now feasible to study them as separate worlds even without the aid of mutual events ({\it e.g. Brown and Calvin}, 2000; {\it Buie and Grundy}, 2000).   Just as the discovery of Charon propelled Pluto science forward, the recent study of two additional moons of Pluto ({\it Weaver et al.}, 2006) can be expected to give another boost to Pluto science, by enabling detailed studies of the dynamics of the system ({\it Buie et al.}, 2006; {\it Lee and Peale}, 2006), and providing new constraints on formation scenarios ({\it e.g. Canup}, 2005; {\it Ward and Canup}, 2006).

\bigskip
\noindent
\textbf{ 1.2 Discovery of Binaries}
\bigskip

The serendipitous discovery of the second TNB, 1998 WW$_{31}$ ({\it Veillet et al.}, 2001) marked a breakthrough for binaries in the Kuiper Belt.  It immediately provided a context for Pluto/Charon as a member of a group of similar systems rather than as a unique oddity.  The relatively large separation and size of the secondary dispelled the notion that Kuiper Belt satellites would all be small, faint, and difficult-to-resolve collision fragments.  Some, at least, were detectable from the ground with moderately good observing conditions as had been foreseen by {\it Toth} (1999).  The next two discovered TNBs were just such systems ({\it Elliot et al.}, 2001; {\it Kavelaars et al.}, 2001).

The first conscious search for satellites of transneptunian objects (TNOs) was carried out by M.~Brown and C.~Trujillo using the Space Telescope Imaging Spectrograph (STIS) on the Hubble Space Telescope (HST) starting in August 2000 ({\it Trujillo and Brown}, 2002; {\it Brown and Trujillo}, 2002).  A series of  large surveys with HST followed producing the discovery of most of the known TNBs (e.g. {\it Noll et al.}, 2002a,b,c, 2003; {\it Stephens et al.}, 2004; {\it Stephens and Noll}, 2006; {\it Noll et al.}, 2006a,b,c,d, e,f; Figure 1).   

Large ground-based surveys -- Keck ({\it Schaller and Brown}, 2003), Deep Ecliptic Survey (DES) followup at Magellan ({\it Millis et al.}, 2002; {\it Ellliot et al.}, 2005 ), and the Canada-France Ecliptic Plane Survey (CFEPS; {\it Allen et al.}, 2006) -- have produced a few detections.  Though significantly less productive because of the limited angular resolution possible from the ground compared to HST, the sheer number of objects observed by these surveys makes them a valuable statistical resource ({\it Kern and Elliot}, 2006).  Both space- and ground-based discoveries are described in detail in \S 2.2.

Binaries may also be ``discovered'' theoretically.  {\it Agnor and Hamilton} (2006) have shown that the most likely explanation for the origin of Neptune's retrograde satellite Triton is the capture of one component of a binary that encountered the giant planet.  The viability of this model is enabled by the paradigm-shifting realization that binaries in the transneptunian region are common.

\bigskip

\section{\textbf{INVENTORY}}

Much can be learned about binaries and the environment in which they formed from simple accounting.  The fraction of binaries in the transneptunian population is far higher than anyone guessed a decade ago when none were yet recognized (except for Pluto/Charon), and considerably higher than was thought even four years ago as the first spate of discoveries was being made.  As the number of binaries has climbed, it has become apparent that stating the fraction of binaries is not a simple task.  The fraction in a given sample is strongly dependent on a number of observational factors, chiefly angular resolution and sensitivity.   To add to the complexity, TNOs can be divided into dynamical groups with possibly differing binary fractions.  Thus, the fraction of binaries in a particular sample also depends on the mix of dynamical classes in the sample.  Perversely, perhaps, the brightest TNOs, and thus the first to be sampled, belong to dynamical classes with a lower overall fraction of sizable binary companions.  Other dependencies may also help determine the fraction and nature of binaries and multiples; for example, very small, possibly collision-produced companions appear to be most likely around the largest of the TNOs.  The current inventory of TNBs, while impressive compared to just a few years ago, remains inadequate to address all of the questions one would like to ask.

\bigskip
\noindent
\textbf{ 2.1 Current Inventory of Transneptunian Binaries}
\bigskip

As of February 2007, more than 40 TNO and Centaur binaries had been announced through the International Astronomical Union Circulars (IAUC) and/or other publications.  Additional binaries are not yet documented in a publication.  All of the binaries of which we are aware are compiled in Table 1.  References listed are generally the discovery announcement, when available.  Values for separation and relative magnitude were recalculated for many of the objects and supersede earlier published values.  The osculating heliocentric orbital parameters $a_{\odot}$, $e_{\odot}$, and $i_{\odot}$ are listed as well as the dynamical class.  For the latter we have followed the DES convention ({\it Elliot et al.}, 2005), with the resonant objects identified by their specific resonance as $n$:$m$, where $n$ refers to the mean motion of Neptune.  Classical objects on orbits of low inclination and eccentricity are designated $C$.  Classical objects with an integrated average inclination $i > 5^{\circ}$ relative to the invariable plane are denoted by $H$.  Objects in the scattered disk are labelled $S$ or $X$ depending on whether their Tisserand parameter, $T$, is less than or greater than 3.  The Tisserand parameter relative to Neptune is defined as $T_N = a_N/a\ + 2[(1-e^2)a/a_N]^{1/2}\ cos(i)$ where $a$,$e$, and $i$ are the heliocentric semimajor axis, eccentricity, and inclination of the TNO and $a_N$ is the semimajor axis of Neptune.  Objects in the extended scattered disk, $X$, have the additional requirement of a time-averaged eccentricity greater than 0.2.    Centaurs and Centaur-like\footnote{Centaur-like objects are in unstable, non-resonant, giant-planet-crossing orbits just like the Centaurs, but have a semimajor axis greater than 30.1 AU.  There is currently disagreement on what this class of objects should be called.  Because of their similarly unstable orbits, we follow the DES convention and group them with Centaurs.} objects, labelled \textcent , are on unstable, non-resonant, planet-crossing orbits and are, therefore, dynamically young.   In Table 1 we list the objects in three broad dynamical groupings, Classical, Scattered, and Resonant.  The Classical grouping includes all Classical objects regardless of inclination, {\it i.e.} both Hot and Cold Classicals.  The Resonant group includes all objects verified to be in mean motion resonances by numerical integration.  The Scattered group includes both Near and Extended Scattered objects and the Centaurs.   Within each group we have ordered the objects by absolute magnitude, $H_V$.

In addition to the dynamical class and osculating heliocentric orbital elements of the binaries, we list in Table 1 three additional measurements available for all of the known binaries: 1.) The reported separation at discovery (in arcsec) is shown with the error in the final significant digit in parentheses.  Separations reported without an error estimate are shown in italics.  As we discuss in more detail below in \S {3.3}, the separation at discovery is not an intrinsic property of binary orbits, but can be useful for estimating the distribution of binary semimajor axes.  2.) The magnitude difference, $\Delta_{\rm mag}$, can be used to derive the size ratio of the components (with the customary assumption of equal albedos).  Once again, errors in the final digit are shown in  parentheses, and estimated quantities are in italics.  3.) The absolute magnitude, $H_V$, is taken from the Minor Planet Center (MPC) and applies to the combined light of the unresolved binary.  Better measurements of  $H_V$ are available for some objects ({\it Romanishin and Tegler}, 2005), but for the sake of uniformity we use the MPC values for all objects.  $H_V$ can provide a determination of the size if the albedo is known or can be estimated.  However, the range of albedo in the transneptunian population is large ({\it Grundy et al.}, 2005) as is the phase behavior ({\it Rabinowitz et al.}, 2007) making any such estimate risky (see also the chapters by {\it Stansberry et al.} and {\it Belskaya et al.}).

\bigskip
\noindent
\textbf{ 2.2 Large Surveys, Observational Limits, and Bias}
\bigskip

Several large surveys have produced the discovery of large fractions of the known TNBs.   Observational limits are, to first order, a function of the telescope and instrument used for the observations.  This is more easily characterized for space-based instruments than for ground-based surveys, but approximate limits for the latter can be estimated.

The largest semi-uniform ground-based survey of TNOs that has been systematically searched for binaries is the Deep Ecliptic Survey ({\it Millis et al.}, 2002; {\it Elliot et al.}, 2005).  {\it Kern and Elliot} (2006) searched 634 unique objects from the DES survey and identified 1.  These observations were made with 4m telescopes at CTIO and KPNO utilizing wide field Mosaic cameras ({\it Muller et al.}, 1998) with 0.5 arcsec pixels.  Median seeing for the entire data set was 1.65 arcsec ({\it Kern}, 2005) and effectively sets the detection limit on angular separation.  Magnitude limits in the broad $V$$R$ filter for a well-separated secondary vary, depending on seeing, from $V$$R$ = 23 to $V$$R$ = 24. 

Follow-up observations of 212 DES objects made with the Magellan telescopes to improve astrometry were also searched for undetected binaries ({\it Kern and Elliot}, 2006).  Most observations were made with the MagIC camera ({\it Osip et al.}, 2004) which has a pixel scale of 0.069 arcsec.  {\it Kern} (2005) reports the median seeing for these observations was 0.7 arcsec with a magnitude limit similar to the DES survey.  Of the 212 objects observed with Magellan, 3 were found to be binaries ({\it Osip et al.}, 2003, {\it Kern and Elliot}, 2005, 2006).

The Keck telescope survey reported by {\it Schaller and Brown} (2003) observed 150 objects and found no new binaries.  The observational limits for this survey have not been published; they are probably similar to the DES as the primary limiting factor is seeing.  

The target lists for the three large ground-based surveys have not been published and it is unclear how much overlap there may be.  However, even duplicate observations of an individual target can be useful since some TNBs are known to have significantly eccentric or edge-on orbits and are, therefore, variable in their detectability.  The one firm conclusion, however, that can be reached from these data is that binaries separated sufficiently for detection with uncorrected ground-based observations are uncommon, occurring around 1--2\% of TNOs at most.  

The most productive tool for finding TNBs is the HST which has found 41 of the 52 companions listed in Table 1.  The combination of high angular resolution, high sensitivity, and stable point-spread-function (PSF) make it ideally matched to the requirements for finding and studying TNBs.  The first conscious search for TNBs using HST was carried out in August 2000--January 2001 in a program that looked at just 2 TNOs and found no companions.  Two other small programs executed between October 1997--September 1998 observed 8 TNOs with the potential to identify a binary, had there been one among the objects observed.  

The first moderate-scale program to search for binaries used STIS in imaging mode to search for binaries around 25 TNOs from August 2001--August 2002.  This program found 2 binaries, both relatively faint companions to the resonant TNOs (47171) 1999 TC$_{36}$ and (26308) 1998 SM$_{165}$ ({\it Trujillo and Brown}, 2002; {\it Brown and Trujillo}, 2002).  The STIS imaging mode has a pixel scale of 50 milliarcsec making it possible to directly resolve objects separated by approximately 100 milliarcsec or more.  In principle, PSF analysis can detect binaries at significantly smaller separations in HST data because of the stability of the PSF ({\it Stephens and Noll}, 2006).  The STIS data were obtained without moving the telescope to track the target's motion.  The TNOs observed this way drift measurably during exposures, complicating the PSF analysis for these data.  

From July 2001--June 2002 75 separate TNOs were observed with WFPC2 in a program designed to obtain V, R, and I band colors ({\it Stephens et al.}, 2003); 3 of these were found to be binary.  To achieve better sensitivity and because of the relatively large uncertainties in TNO orbits at the time, the targets were observed with the WF camera with 100 milliarcsec pixels.  The sensitivity to faint companions for these data has not been fully quantified and exposure times were varied depending on the anticipated brightness of the TNO so that, in any case, the limits vary.  Typical photometric uncertainties ranged from 3--8\% for V magnitudes that ranged from 21.9 to 25.1 with a median magnitude of 23.6.  

NICMOS was used to observe 82 separate TNOs from August 2002 through June 2003.  Observations were made with the NIC2 camera at a 75 milliarcsec pixel scale.  Two broad filters, the F110W and F160W, approximating the J and H band filters, were used in the observations.  A total of 9 new binary systems were identified from this data set, 3 of which were resolved and visible in the unprocessed data.  The other 6 binaries were identified from a process of PSF fitting that enabled a significant increase in detectivity beyond the usual Nyquist limit with a separation/relative brightness/total brightness function that is complex and modeled with simulated binaries.  Several of the binaries identified in this way have been subsequently resolved with the higher resolution HRC, verifying the accuracy of the analysis of the NICMOS data ({\it Stephens and Noll}, 2006).

From July 2005 through January 2007, HST's HRC was used to look at more than 100 TNOs.  This program identified a significant number of new binaries ({\it Noll et al.}, 2006 a,b,c,d,e,f).  The pixels of the HRC, while operational, were a distorted 25 x 28 milliarcsec quadrilateral.  Observations with the Clear filter were able to reach a faint limiting magnitude of at least V=27, significantly deeper than most other HST observations of TNOs ({\it Noll et al.}, 2006h).

The wide variety of instruments and inherent observing limitations for each makes analysis from a concatenation of these data extremely problematic.  Indeed, even observations taken with the same instrument vary in their limits based on exposure time, focus, seeing, and other systematics.  This variability is reduced for space-based observations, but not entirely eliminated.  Additionally, the detection of close pairs depends on the ability to model and subtract the PSF of the primary with a resulting spatially-dependent detection limit.

\bigskip
\noindent
\textbf{ 2.3 Binary Frequency}
\bigskip

Does the fraction of TNBs vary with any of the observable properties of TNOs?  In the search for correlations, the most promising quantities are those that can be associated with formation or survival.

Dynamical class is one such quantity because objects in different dynamical classes may have had different origins and dynamical histories.  On one extreme, objects on unstable, non-resonant, planet-crossing orbits (the Centaurs and similar objects) have many close encounters with giant planets during their lifetimes. These close encounters can potentially disrupt weakly bound binaries ({\it Noll et al.}, 2006g; {\it Petit and Mousis}, 2004; see also \S3.3).  At the other extreme, objects in the classical disk may have persisted largely undisturbed from the protoplanetary disk and thus be a more congenial environment for survival of binaries.  {\it Stephens and Noll} (2006) have shown that, indeed, the binary frequency in the cold classical disk is significantly higher than for other dynamical classes at the observational limits attainable by the NIC2 camera.  They found that 22$\pm {10\atop 5}$\% of classical TNOs with inclinations less than 5$^{\circ}$ had detectable binaries at magnitude-dependent separations typically $>$ 60 milliarcsec.  The rate of binaries for all other dynamical classes combined was  5.5$\pm {4\atop 2}$\% for the same limits.  The number of objects and binaries in this dataset precluded any meaningful conclusions on the binary frequency in the various dynamically ``hot'' populations.  Including more recent HRC observations strengthens these conclusions, as shown in Figure 2.

The size of the primary is another parameter that could, potentially, be correlated with binary status.  To the extent that size correlates with dynamical class, the two sorting criteria can be confused for one another.  There are, in fact, differences in size distributions as a function of dynamical class with the resonant objects, scattered disk, and high inclination classical objects all having significantly larger upper size limits than the cold classical population ({\it Levison and Stern}, 2001; {\it Bernstein et al.}, 2004;).  The published binary searches to date cover an insufficient range in size (as measured by $H_V$) to be able to reach any strong conclusions with regard to binary frequency as a function of size.  {\it Brown et al.} (2006a) have noted an apparently higher fraction of bound systems among the largest TNOs based on the satellites known for Pluto, Eris, and 2003 EL$_{61}$.  With the size bins chosen in this work (the four largest TNOs versus smaller TNOs) the difference in binary frequency is mathematically significant.  The subsequent detection of small satellites around four more large TNOs ({\it Brown and Suer}, 2007) strengthens this apparent trend.  In considering whether small satellites exist around smaller TNOs, it is important to consider that companions as faint as those of the largest TNOs would only be detectable for widely separated companions in a subset of the most recent deep observations (Figure 3 {\it Noll et al.}, 2006g) and would not have been detectable in most earlier surveys.  At the same fractional Hill radius, small satellites of 100-km class TNOs would be extremely difficult to detect in any existing observations.  Caveats aside, however, it seems reasonable to hypothesize that the small satellites of the largest TNOs may be collisional in origin ({\it Brown et al.}, 2006a; {\it Stern et al.}, 2006) while the nearly equal sized binaries of smaller TNOs may form from dynamical capture and thus, there may be real differences in the frequencies of these two types of bound systems.  We discuss these two modes of formation in more detail in \S {4}.

An obvious, but sometimes neglected, qualifier that must accompany any description of binary frequency is the limit in magnitude and separation imposed by the observational method used to obtain the data.  This is especially true because, as discussed in more detail below and as shown in Figure 4, the number of binaries detectable in a given sample appears to be a strong function of separation ({\it Kern and Elliot}, 2006).  The variation of binary frequency with dynamical class and/or size means that statements about binary frequency must be further qualified by a description of the sample.  Thus, it is impossible to state a unique ``frequency of transneptunian binaries''.  The need for a more nuanced description of binary frequency among the transneptunian populations is both a challenge and an opportunity waiting to be exploited.

\section{\textbf{PHYSICAL PARAMETERS}}

\bigskip
\noindent
\textbf{ 3.1 Relative Sizes of Binary Components}
\bigskip

The relative sizes of the primary and secondary components is an important physical parameter that is usually available from a single set of observations (with the important caveat of possible non-spherical shapes and lightcurve variations as we discuss in \S {3.6}).  The observed magnitude difference between the two components of a binary, $\Delta_{\rm mag}$, can be used to constrain the ratios of their radii $R_1/R_2$ and surface areas $A_1/A_2$ according to

\begin{equation}
\frac{R_1}{R_2} = \sqrt{\frac{A_1}{A_2}} =
\sqrt{\frac{p_2}{p_1}} 10^{0.2\Delta_{\rm mag}} \,.
\label{eqn:relsize}
\end{equation}

\bigskip

\begin{deluxetable}{llllllllr}
\tabletypesize{\small}
\tablecaption{Transneptunian Binaries}
\tablewidth{0pt}
\tablehead{                               & dyn.    &\multispan3{heliocentric orbit elements} &separation&                       &             &     \\ 
\phantom{(123456)} object     & class   &  $a_{\odot}$ (AU) &  $e_{\odot}$ &\ $i_{\odot}$ ($^\circ$) &$s_0$ (arcsec)    &$\Delta_{\rm mag}$ &H$_V$ &ref \phantom{xxx} \\ }
\startdata
\smallskip
{\bf Classical} &&&&&&&&\\
\ \ (50000) Quaoar                                      & H   & 43.609& 0.037   &  8.0   & 0.35(1)   & 5.6(2)       &\ 2.6  & [1]       \\ 
\ \ (79360) 1997 CS$_{29}$                      & C  & 43.876 & 0.013  & 2.2   & 0.07(1)     & 0.09(9)       & \ 5.1 & [2]    \\ 
(148780) 2001 UQ$_{18}$                       & C  & 44.545 & 0.057  & 5.2  & 0.177(7)     & 0.7(2)        & \ 5.1 & [$\dagger$]     \\ 
\phantom{(123456)} 2003 QA$_{91}$    & C  & 44.157 & 0.067  & 2.4  & 0.056(4)     & 0.1(6)        & \ 5.3 & [$\dagger$]     \\ 
\phantom{(123456)} 2001 QY$_{297}$  & C  & 43.671 & 0.081  & 1.5  & 0.091(2)     & 0.42(7)      & \ 5.4 & [$\dagger$]     \\ 
\ \ (88611) 2001 QT$_{297}$                   & C  & 44.028 & 0.028   & 2.6  & 0.61(2)      & 0.7(2)         & \ 5.5 & [3]    \\ 
\phantom{(123456)} 2001 XR$_{254}$  & C  & 43.316 & 0.023   & 1.2  & 0.107(2)    & 0.09(6)      & \ 5.6 & [$\dagger$]     \\ 
\phantom{(123456)} 2003 WU$_{188}$  & C  & 44.347 & 0.039  & 3.8  & 0.042(4)    & 0.7(3)        & \ 5.8 & [$\dagger$]     \\ 
\ \ (66652)  1999 RZ$_{253}$                  	& C  & 42.779 & 0.090   & 0.6  & 0.21(2)     & 0.33(6)       & \ 5.9 & [4]    \\ 
(134860) 2000 OJ$_{67}$                       	& C  & 42.840 & 0.023  & 1.1   & 0.08(1)     &{\it 0.8}      & \ 6.0 &  [2]   \\
\phantom{(123456)} 2001 RZ$_{143}$    & C  & 44.282 & 0.068  & 2.1  & 0.046(3)   & 0.1(3)         & \ 6.0 & [$\dagger$]     \\ 
\phantom{(123456)} 1998 WW$_{31}$ 	& C  & 44.485 & 0.089   & 6.8  & {\it 1.2}   &{\it 0.4}       & \ 6.1 & [5]    \\ 
\phantom{(123456)} 2005 EO$_{304}$ 	& C  & 45.966 & 0.080  & 3.4   & 2.67(6)     & 1.2(1)         & \ 6.2 & [6]    \\ 
\phantom{(123456)} 2003 QR$_{91}$    & H  & 46.361 & 0.183  & 3.5  & 0.062(2)     & 0.2(3)        & \ 6.2 & [$\dagger$]     \\ 
\ \ (80806)  2000 CM$_{105}$      	& C  & 42.236 & 0.064  & 6.7   & 0.059(3)   & 0.6(1)         & \ 6.3 & [2]    \\
\phantom{(123456)} 2003 QY$_{90}$   	& C  & 42.745 & 0.052  & 3.8  & 0.34(2)      & 0.1(2)$^*$  & \ 6.3 & [7]     \\ 
(123509) 2000 WK$_{183}$                     & C & 44.256 & 0.044  & 2.0   & 0.080(4)   & 0.4(7)          &  \ 6.4 & [8]   \\ 
\ \ (58534)  Logos/Zoe                               	& C  & 45.356 & 0.119  & 2.9  & 0.20(3)      & 0.4(1)$^*$ & \ 6.6 & [9]    \\ 
\phantom{(123456)} 2000 CQ$_{114}$  	& C  & 46.230 & 0.110  & 2.7  & 0.178(5)    & 0.4(2)         & \ 6.6 & [10]    \\ 
\phantom{(123456)} 2000 CF$_{105}$ 	& C  & 43.881 & 0.037  & 0.5  & 0.78(3)      & 0.7(2)         & \ 6.9 & [11]  \\ 
\phantom{(123456)} 1999 OJ$_4$         	& C  & 38.067 & 0.023  & 4.0  &0.097(4)     & 0.16(9)       & \ 7.0 & [2]    \\ 
\phantom{(123456)} 2001 FL$_{185}$   & C  & 44.178 & 0.077  & 3.6  & 0.065(14)   & 0.8(6)        & \ 7.0 & [$\dagger$]     \\ 
\phantom{(123456)} 2003 UN$_{284}$  	& C  & 42.453 & 0.010  & 3.1  & 2.0(1)        & 0.6(2)         & \ 7.4 & [12]   \\ 
\phantom{(123456)} 2001 QW$_{322}$ 	& C  & 44.067 & 0.027  & 4.8  &{\it 4}        & 0.0(1)         & \ 7.8 & [13]  \\ 
\phantom{(123456)} 1999 RT$_{214}$   	& C  & 42.711 & 0.052  & 2.6  & 0.107(4)    & 0.81(9)       & \ 7.8 & [14]  \\ \bigskip 
\phantom{(123456)} 2003 TJ$_{58}$   	& C  & 44.575& 0.089  & 1.0  & 0.119(2)     & 0.50(7)       & \ 7.8 & [$\dagger$]     \\ 
%
\smallskip
{\bf Scattered} &&&&&&&&\\
(136199) Eris                                           	& S   & 67.695 & 0.441   & 44.2  &0.53(1)    & 4.43(5)       & -1.2 & [15]   \\ 
(136108) 2003 EL$_{61}$                    	& S   & 43.316 & 0.190   & 28.2  & 0.63(2)   & 3.1(1)         & \ 0.2 & [16]  \\ 
                                                                   &      &             &              &          & 0.52(3)   & 4.6(4)         &         & [17]  \\ 
\ \  (55637)   2002 UX$_{25}$                  & S  & 42.551  & 0.141   & 19.5 & 0.164(3)  & 2.5(2)         & \ 3.6 & [1]  \\ 
(120347) 2004 SB$_{60}$                     	& S  & 42.032 & 0.104   & 24.0  &  0.107(3) & 2.2(1)         & \ 4.4 & [18]  \\ 
\ \ (48639)  1995 TL$_8$                          	& X  & 52.267 & 0.235   & 0.2   & 0.01(1)    & {\it 1.7}      & \ 5.4 & [2]    \\ 
\phantom{(123456)} 2001 QC$_{298}$	& S  & 46.222 & 0.123   & 30.6  & 0.130(7)  & 0.58(3)       & \ 6.1 & [19]  \\  
\phantom{(123456)} 2004 PB$_{108}$	& S  & 44.791 & 0.096   & 20.3  & 0.172( 3) & 1.2(1)         & \ 6.3 & [20]  \\  
\ \ (65489)  {Ceto/Phorcys}             & \textcent &102.876 & 0.821  & 22.3  & 0.085(2)  & 0.6(1)         & \ 6.3 & [21] \\ %
\phantom{(123456)} 2002 GZ$_{31}$	& X  & 50.227 & 0.237   & 1.1   & 0.070(9)  & 1.0(2)         & \ 6.5 & [22]  \\ 
\ \ (60458)  2000 CM$_{114}$                  & S  & 59.838  & 0.407  & 19.7  &  0.074(6) & 0.57(7)       & \ 7.0 & [23]   \\ \bigskip 
\ \ (42355)  {Typhon/Echidna}        & \textcent & 38.112  & 0.540  & 2.4    & 0.109(2)  &1.47(4)        & \ 7.2 & [24] \\ 
%
%
\smallskip
{\bf Resonant} &&&&&&&&\\
(134340)  Pluto/Charon                              & 3:2 & 39.482 & 0.248    & 17.1  &{\it 0.9}  &{\it 2--3}    & -1.0 & [25]  \\ 
\phantom{(134340)  Pluto/}Nix                 &       &             &              &          &1.85(4)    & 9.26(2)     &         & [26]  \\ 
\phantom{(134340)  Pluto/}Hydra             &       &             &              &          &2.09(4)    & 8.65(2)     &         & [26]  \\ 
\ \ (90482) Orcus                                         & 3:2 &39.386  & 0.220    & 20.6 &0.256(2)  & {\it 2.5}    &\ 2.3  & [1]       \\  
\phantom{(123456)} 2003 AZ$_{84}$      & 3:2 & 39.414 & 0.181    & 13.6 &0.22(1)   & 5.0(3)        & \ 3.9 & [1] \\  %
\ \  (47171)   1999 TC$_{36}$                    & 3:2 & 39.270 & 0.222   & 8.4    &0.367(4)  & 2.21(1)     & \ 4.9 & [27]  \\ 
\ \ (82075)  2000 YW$_{134}$                  & 8:3 & 57.779 & 0.287    & 19.8  & 0.06(1)   & {\it 1.3}    & \ 5.0 & [2]     \\
(119979)  2002 WC$_{19}$                   	 & 2:1 & 47.625 & 0.260   & 9.2    & 0.090(8) &  2.5(4)       & \ 5.1 & [28]\\ 
\ \ (26308)  1998 SM$_{165}$                   & 2:1 & 47.501 & 0.370   & 13.5  & 0.205(1)  & 2.6(3)       & \ 5.8 & [29]\\ 
\phantom{(123456)} 2003 QW$_{111}$ 	 & 7:4 & 43.659 & 0.111   & 2.7    &  0.321(3) & 1.47(8)     & \ 6.2 & [30] \\  
\phantom{(123456)} 2000 QL$_{251}$    & 2:1 & 47.650 & 0.216   & 3.7    & 0.25(6)   & 0.05(5)      & \ 6.3 & [31] \\  
\ \  (60621)   2000 FE$_{8}$                      & 5:2 & 55.633 & 0.404   & 5.9    & 0.044(3)  & 0.6(3)       & \ 6.7 & [20]  \\ \bigskip%
(139775)  2002 QG$_{298}$                   	& 3:2 & 39.298 & 0.192   & 6.5   &{\it contact?} & {\it N/A}& \ 7.0 & [32]\\ 
\enddata

\tablecomments {Objects are sorted into three dynamical groups, Classical (both hot, $H$, and cold $C$), Scattered (includes Scattered-near, $S$, Scattered-extended, $X$, and Centaurs, \textcent) and Resonant.  Objects in each grouping are sorted by absolute magnitude, $H_V$.  Uncertainties in the last digit(s) of measured quantities appear in parentheses.  Table entries in italics indicate quantities that have been published without error estimates or that have been computed by the authors from estimated quantities.  The $H_V$ column lists the combined absolute magnitude of the system as tabulated by the Minor Planet Center (MPC).  $^*$ The brightness of Logos varies significantly as described in \S {3.6}.  References: [1] {\it Brown and Suer} (2007); [2] {\it Stephens and Noll} (2006); [3] {\it Elliot et al.}\ (2001); [4] {\it Noll et al.}\ (2003); [5] {\it Veillet et al.}\ (2001); [6] {\it  Kern and Elliot} (2005); [7] {\it Elliot, Kern, and Clancy} (2003); [8] {\it Noll et al.}\ (2007a); [9] {\it Noll et al.}\ (2002a); [10] {\it Stephens, Noll, and Grundy} (2004); [11] {\it Noll et al.}\ (2002b); [12] {\it Millis and Clancy} (2003); [13] {\it Kavelaars et al.}\ (2001); [14] {\it Noll et al.}\ (2006f); [15] {\it Brown} (2005a); [16] {\it Brown et al.}\ (2005b); [17] {\it Brown} (2005b); [18] {\it Noll et al.}\ (2006e); [19] {\it Noll et al.}\ (2002c); [20] {\it Noll et al.}\ (2007d); [21] {\it Noll et al.}\ (2006g), {\it Grundy et al.}\ (2006); [22] {\it Noll et al.}\ (2007c); [23] {\it Noll et al.}\ (2006b); [24] {\it Noll et al.}\ (2006a); [25] {\it Smith et al.}\ (1978), {\it Christy and Harrington} (1978);  [26] {\it Weaver et al.}\ (2005); [27] {\it Trujillo and Brown} (2002); [28] {\it Noll et al.}\ (2007b); [29] {\it Brown and Trujillo} (2002); [30] {\it Noll et al.}\ (2006c); [31] {\it Noll et al.}\ (2006d);  [32] {\it Sheppard and Jewitt} (2004);[$\dagger$] unpublished as of 28 February 2007.
}

\end{deluxetable}

%


\begin{figure*}
 \epsscale{1.5}
 \plotone{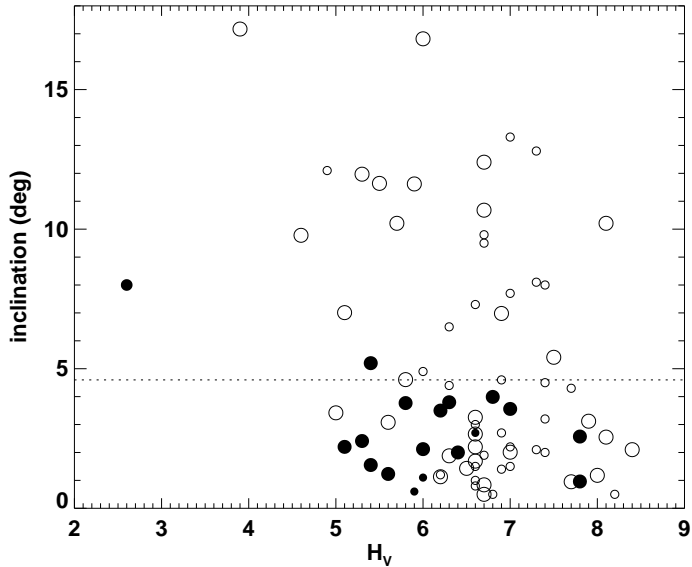}
 \caption{\small Classical TNOs observed by HST are plotted with their absolute magnitude, $H_V$, on the horizontal axis and their inclination to the ecliptic, $i$, plotted vertically.  Objects observed with the HRC (~30 milliarcsec resolution) are shown as large circles; less sensitive observations made with NIC2 (~75 milliarcsec resolution) are shown as smaller circles.  Binaries are shown as filled circles.  The dotted line at 4.6$^{\circ}$ is the boundary between Hot and Cold Classical populations proposed by {\it Elliot et al.} (2005).  The extremely strong preference for low inclination binaries in this sample is evident. }  
 \end{figure*}

\noindent where $p_1$ and $p_2$ are the albedos of the TNB components.  It is usual, but not necessary, to assume that both components have the same albedo.  As we note  in \S {3.6}, the similarity of the colors of TNB components suggests common surface materials with similar albedos may be the norm for TNBs.  However, we also note that in the one instance where separate albedos have been measured, the Pluto/Charon system, they are not identical.  Given the large range of albedos of TNOs of all sizes ({\it e.g.~Grundy et al.}, 2005; chapter by {\it Stansberry et al.}), it is reasonable to keep this customary simplification in mind.

The relative sizes of TNBs found so far is heavily skewed to nearly-equal sized systems as can be seen in Table 1 and Figure 3.  The prevalence of nearly equal-sized systems is a unique feature of TNBs compared to binaries in the Main Belt or Near Earth populations ({\it e.g.~Richardson and  Walsh}, 2006; {\it Noll}, 2006).  To some extent, this conclusion is limited by observational bias, but, deep surveys with the HST's HRC show an apparent lack of asymmetric binaries ({\it Noll et al.}, 2006h; Figure 3).  A preference for similar sized components is a natural outcome of dynamical capture models for the formation of binaries ({\it Astakhov et al.}, 2005) and may explain the observed distribution of relative sizes after accounting for the small number of large object satellites that appear to have been formed from collisions ({\it e.g.~Stern et al.}, 2006).

\begin{figure*}
 \epsscale{1.3}
 \plotone{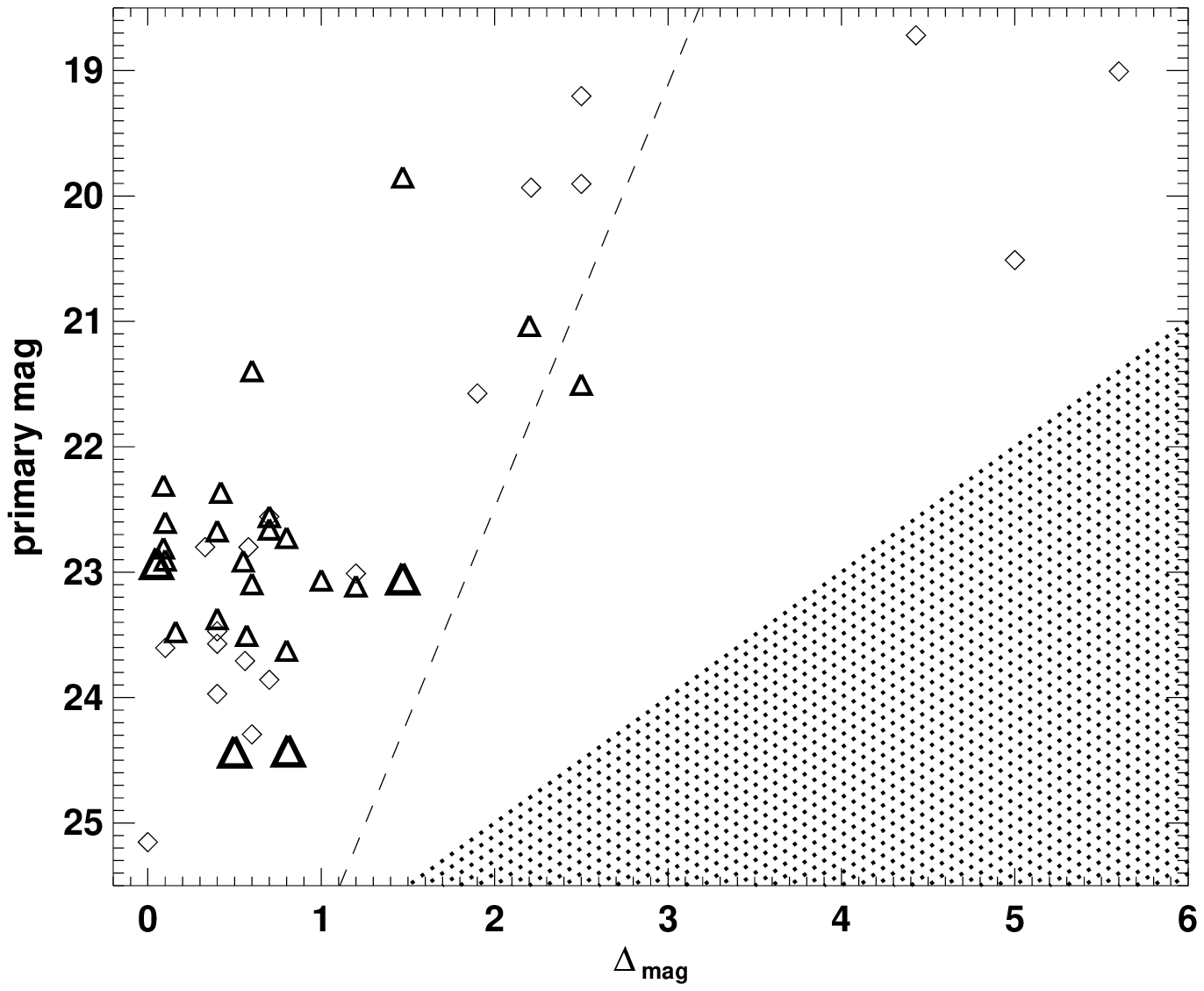}
 \caption{\small In this figure the y-axis is the observed magnitude of the primary component of observed TNBs.  The x-axis shows the range of observed and detectable magnitude differences, $\Delta_{\rm mag}$, for secondaries under several different observing circumstances.  Binaries detected with HST's HRC are shown as triangles, binaries found with other instruments are shown as diamonds.  The stippled area shows the 7-sigma detection limit for companions that are sufficiently separated that their local background is dominated by sky noise and dark current.  Objects that are observed by the HRC and are background dominated are shown as large triangles.  The background and detection limits for the objects shown by the small triangles are limited to a varying degree by the PSF of the primary.  The dashed line defines an approximate empirical detection limit for objects separated by 3-pixels from their primary, {\it i.e.}~75 milliarcsec.  The background at this separation is dominated by the PSF of the primary to a degree that varies as a function of the primary.  Both detection limits are dependent on the details of the observations and therefore do not apply to the observations from other instruments (diamonds) which are less sensitive.  The clustering of binaries at $\Delta_{\rm mag}< 1$ appears to be an intrinsic property of TNBs and not an observational bias.  (Figure from {\it Noll et al.}, 2006h) }  
 \end{figure*}

\bigskip

\bigskip
\noindent
\textbf{ 3.2 Orbit Determination}
\bigskip

Determining the mutual orbit of a spatially-resolved pair of gravitationally bound objects is a classic problem of celestial mechanics.  Solutions are reviewed in numerous textbooks  ({\it e.g. Aitken}, 1964; {\it Heintz}, 1978; {\it Smart}, 1980; {\it Danby}, 1992).

The Keplerian orbit of a pair of bound point masses can be described by seven independent quantities.  There is considerable flexibility in the choice of the seven.  One possible set is the system mass plus the three dimensional relative position and velocity vectors at a specific time.  This form is generally preferred for specifying inputs to orbital integration routines.  A second set of elements is more frequently used to specify binary orbits in the scientific literature: period $T$, semimajor axis $a$, eccentricity $e$, inclination $i$, mean longitude at a specific time $\epsilon$, longitude of the ascending node $\Omega$, and longitude of periapsis $\varpi$\null.  Other sets of elements are possible and are occasionally encountered in the literature.

Each observation of a resolved binary pair provides two constraints, the separation and position angle, or equivalently, the relative positions in right ascension and  declination.  To constrain seven orbital elements requires at least four observations, in order to have more constraints than unknowns.  In practice, four observations are often not enough to uniquely determine the seven unknowns, because the constraints from the separate observations are not necessarily independent of one another.  For example, four observations that happen to sample the same orbital longitude do little to constrain the shape and orientation of the orbit.  Many more than four observations may be needed, if they are poorly timed.

Instruments capable of doing relative astrometry on the extremely close-spaced and faint components of TNBs are a scarce and valuable resource.  To make the most efficient use of these facilities requires strategic timing of follow-up observations to maximize the additional constraints provided by each successive observation.  Several groups have applied Monte Carlo techniques to this task ({\it e.g., Margot et al.}, 2004, 2005; {\it Hestroffer et al.} 2005; {\it Grundy et al.} 2007).  The general approach is to generate random orbits consistent with the existing set of observations which do not yet uniquely determine the binary orbit.  The collection of orbits produced by this exercise is used to map out regions of orbital element space consistent with the constraints already available, and to identify times when follow-up observations would be most effective for collapsing the cloud of possible orbits.

An additional complication involves near-equal brightness binaries, which are not uncommon among the TNBs.  When the ``primary'' and ``secondary'' have similar brightness, or large amplitude lightcurve variations, they become difficult to distinguish from one another, leading to uncertainties which require additional observations to resolve.  The binary system 2003 QY$_{90}$ provides a recent example of this situation ({\it Kern and Elliot}, 2006).

Various non-linear least squares techniques such as the Levenburg-Marquardt and AMOEBA algorithms ({\it Press et al.}, 1992) can be used to fit a set of orbital elements to the observational data by iteratively minimizing the residuals (as measured by the $\chi^2$ statistic) between the astrometric data and the positions computed from the orbital elements.  In situations where observational data are particularly abundant, linear least squares fitting techniques may be applicable.  Knowledge of the uncertainties in the fitted orbital elements is just as important as knowledge of the elements themselves.  The Monte Carlo techniques mentioned earlier can be used to directly investigate uncertainties in orbital elements ({\it e.g., Virtanen et al.}, 2001, 2003; {\it Hestroffer et al.}, 2005).  Another approach involves varying each parameter around its best fit value and allowing the remaining parameters to readjust themselves to compensate, resulting in a new fit with different $\chi^2$.  This approach can be used to map out a seven-dimensional $\chi^2$ space, within which contours of 1~$\sigma$ or 3~$\sigma$ likelihood can be traced ({\it Lampton}, 1976; {\it Press et al.}, 1992).

For TNBs, relative motion between the Earth and the orbiting pair leads to changing viewing geometry over time.  The motion has two components: the comparatively rapid motion of the Earth around the sun and the much slower motion of the TNB along its heliocentric orbit.  The resulting changes in viewing geometry complicate the orbit-fitting process compared with procedures developed for binary stars which assume fixed geometry between the observer and the binary system.  However, there are advantages to observing a binary from different angles.  Relative astrometry is otherwise not able to distinguish between the actual orbit and its mirror reflection through the sky plane.  For TNBs, the ambiguity between these two ``mirror'' orbits can be broken by observations spanning as little as a few years ({\it e.g. Hestroffer and Vachier}, 2006), but this has only actually been done for a small minority of orbits to date.

For larger and closer binaries, departures from spherical symmetry of the primary's mass distribution (parametrized by the gravitational harmonic coefficients, primarily $J_2$) can exert torques on the secondary leading to secular precession of the line of nodes and\slash or of the line of apsides ({\it e.g., Brouwer and Clemence}, 1961).  Observations of this secular evolution can be used to measure $J_2$ as has been done for a binary asteroid ({\it Marchis et al.}, 2005), but this has not yet been done for any TNBs except Pluto (and that measurement by {\it Descamps}, 2005 neglected possible perturbations on the orbit of Charon by Pluto's smaller satellites).  Comparable effects can be anticipated for the orbits of the more distant satellites of close pairs, such as the orbits of Hydra and Nix around the Pluto\slash Charon binary ({\it Buie et al.} 2006; {\it Lee and Peale}, 2006).

\begin{deluxetable}{lllllllllr}
\tabletypesize{\small}
\tablecaption{Physical Properties of Transneptunian Binaries}
\tablewidth{0pt}
\tablehead{object                             & \ $a$              & \ $e$           &\  $T$            & \ \ $M$           & $p_{\lambda}$           & \ $\rho$            &  $J/J'$       & refs       \\ 
                                                         & (km)              &                    & (days)           &(10$^{18}$ kg)&                                    & (g cm$^{-3}$) &                  &                 \\ }
\startdata
(136199) Eris                                    & {\it 36,000}  & {\it 0}         & {\it 14}         &{\it 16,400}    & 0.86(7) (V)                   & 2.26(25)           & {\it 0.16}  & [1]     \\ 
(134340) Pluto/Charon                     &                      &                    &                       & 14,570(9)       &                                      &                         & {\it 0.40 } & [2]     \\
\phantom{(134340)\ \ \ \ \ }{\it Pluto}    &                      &                    &                       & 13,050(620)   & 0.51--0.71 (V)       & 2.03(6)             &                & [2]     \\
\phantom{(134340)\ \ \ \ \ }{\it Charon}&  19,571(4)     & 0.00000(7) & 6.387230(1)   & 1,521(65)       & 0.38 (V)                 & 1.65(6)            & {\it  -     }  & [4]    \\              
\phantom{(134340)\ \ \ \ \ }{\it Nix}      &  48,670(120)  & 0.002(2)     & 24.8562(13)   &{\it 0.1--2.7}   &{\it 0.01--0.35}      & {\it 2.0}            & {\it  -     }  & [4]    \\              
\phantom{(134340)\ \ \ \ \ }{\it Hydra}  &  64,780(90)    & 0.005(1)    & 38.2065(14)   &{\it 0.2--4.9}   &{\it 0.01--0.35}       & {\it 2.0}          & {\it  -     }  & [4]    \\              
(136108) 2003 EL$_{61}$                      & 49,500(400)  & 0.050(3)      & 49.12(3)        & 4,210(100)     & 0.7(1) ($v$)               & 2.9(4)              & {\it 0.53}  & [3]    \\
\ \ (47171)   1999 TC$_{36}$                  &7,720(460)    & 0.22(2)        & 50.4(5)          & 14.4(2.5)        & 0.08(3) ($v$)             & 0.5(3/2)          & {\it 0.31}   & [5]    \\
\phantom{(123456)} 2001 QC$_{298}$&3,690(70)       & 0.34(1)        & 19.2(2)          & 10.8(7)           &{\it 0.08} ($V_{606}$)&{\it 1.0}         & {\it 1.16}   & [6]    \\ 
\ \ (26308)  1998 SM$_{165}$                &11,310(110)   & 0.47(1)        & 130(1)           & 6.78(24)         & 0.08(3/2) (V)             & 0.7(3/2)           & {\it 0.56}  & [7]    \\
\ \ (65489)   Ceto/Phorcys                        & 1840(50)       & $<$0.014    &9.557(8)         & 5.4(4)             & 0.08(2) (V)                & 1.4(6/3)           & {\it 0.89}  &[8]     \\
\ \ (66652)  1999 RZ$_{253}$                & 4660(170)     & 0.46(1)        & 46.263(6/74) & 3.7(4)             & {\it  0.17} (R)            & {\it 1.0}          & {\it 1.56}  & [9]     \\
\phantom{(123456)} 1998 WW$_{31}$ & 22,300(800)  & 0.82(5)        & 574(10)        & 2.7(4)             & {\it  0.054} (R)          & {\it 1.0}          & {\it 2.22}  & [10]   \\
\ \ (88611)  2001 QT$_{297}$                & 27,880(150)  & 0.241(2)      & 825(1)           & 2.51(5)           & {\it  0.13} ({\it r'})     & {\it 1.0}         & {\it 3.41}  & [11]    \\
\ \ (42355)   Typhon/Echidna                   & 1830(30)       & 0.53(2)        &18.971(1)       & 0.96(5)           & 0.05(V)                      & 0.47(18/10)    & {\it 2.13}  & [12]    \\
\ \ (58534)  Logos/Zoe                             & 8,010(80)      & 0.45(3)        & 312(3)           & 0.42(2)           & {\it 0.37(4)} (R)         &{\it 1.0}          & {\it 2.65}  & [13]    \\
\phantom{(123456)} 2003 QY$_{90}$   &{\it 7,000--13,000}&{\it 0.44--0.93} &{\it 306--321} &{\it0.3-1.7}                &  {\it 0.19--0.35} (r')                & {\it 1.0}         & {\it 2.14} & [14]\\
\enddata

\tablecomments {Objects are listed in order of decreasing system mass.  Uncertainties in the last digit(s) of measured quantities appear in parentheses.  Table entries in italics indicate quantities that have been published without error estimates or that have been computed by the authors from estimated quantities. The components of the Pluto system are listed separately because they have had independent determinations of radius, mass, albedo, and density that are not yet available for other binary or multiple systems.  The wavebands used to determine geometric albedos are shown in parentheses ($v$ is for references citing ``visual" albedo, $V_{606}$ is the HRC's F606W filter).  Densities of $1.0$ g cm$^{-3}$ are assumed for objects without an independent size measurement.  The corresponding geometric albedos are for this assumed unit density.  References: [1] {\it Brown et al.}\ (2006a,b); [2] {\it Buie et al.}\ (1997), {\it Buie et al.}\ (2006), {\it Rabinowitz et al.}\ (2006); [3] {\it Brown et al.}\ (2005); [4]  {\it Buie and Grundy}\ (2000), {\it Buie et al.}\ (2006), {\it Lee and Peale}\ (2006); [5] {\it Stansberry et al.}\ (2006); [6] {\it Margot et al.}\ (2004); [7] {\it Margot et al.}\ (2004), {\it Spencer et al.}\ (2006); [8] {\it Grundy et al.}\ (2007); [9] {\it Noll et al.}\ (2004a); [10] {\it Veillet et al.}\ (2002); [11] {\it Osip et al.}\ (2003), {\it Kern}\ (2005); [12] {\it Grundy et al., in press}; [13] {\it Noll et al.}\ (2004b), {\it Margot et al.}\ (2004); [14] {\it Kern et al., in preparation}}

\end{deluxetable}

\begin{figure*}
 \epsscale{1.5}
 \plotone{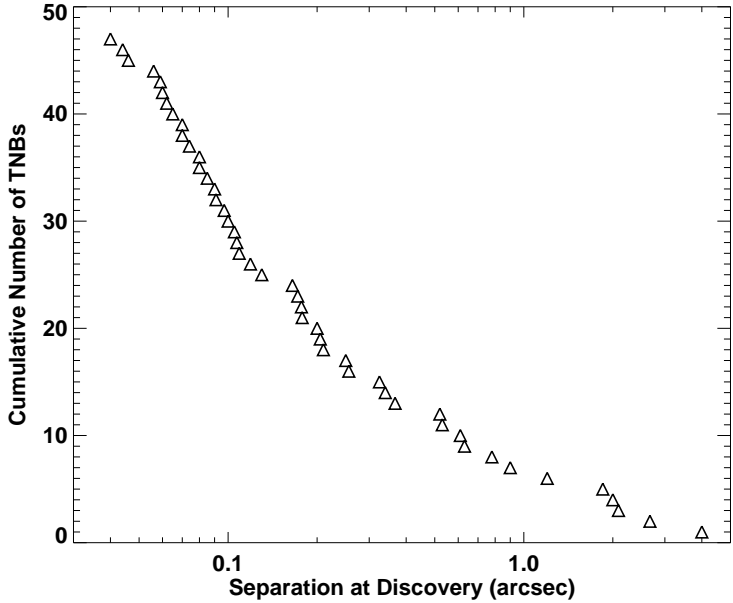}
 \caption{\small The number of objects with a separation greater than or equal to a given separation is shown.  The number of objects at small separation increases faster than an exponential as can be seen in this logarithmic plot.  Observational bias favors detection of widely separated binaries suggesting that the underlying distribution may rise even more rapidly with decreased separation. }  
 \end{figure*}

\bigskip

\bigskip
\noindent
\textbf{ 3.3 Semimajor Axis Distribution}
\bigskip

The separation of components in binaries is a fundamental datum that can be diagnostic in comparing systems and in understanding general processes affecting populations.  Semimajor axes are known for only the subset of TNBs with measured orbits (Table 2).  Unfortunately, from a statistical perspective, the available data are too few and are heavily biased against small semimajor axes by limitations of both discovery and follow-up observations.

As a proxy for the semimajor axis, we can use the separation at discovery, $s_0$, listed in Table 1.  In any individual case this is only an approximate substitute for a measurement of the semimajor axis from an orbit because of the unknown orientation of the orbit plane with respect to the line of sight and the unknown eccentricity of the binary orbit.  Observational biases tend to favor detection of objects close to their widest separation from the primary independent of the orientation of the orbit plane.  For the 13 systems in Table 2 the ratio of the semimajor axis to the distribution of separation at discovery, $a/s_0$, ranges from 0.5 to 2.  The median value of $a/s_0$ is 0.96.  Thus, for a moderate size ensemble, the distribution of separations at discovery provides an excellent statistical approximation for the semimajor axis distribution.  The separations at discovery from Table 1 are plotted in Figure 4.  This log plot shows a stronger than exponential increase at smaller separations.  The trend of increasing binary frequency at small separation is robust because observational bias will decrease the number of detectable companions near the limits of instrument resolution.  Interestingly, this trend is in qualitative agreement with the binary formation model discussed in more detail in \S {4.5}.

It is natural to wonder if the trend of binary frequency continues at even smaller separations or if there is some eventual cutoff.  The separation of TNB components in terms of ``typical'' values for the distance to Earth in AU, $\Delta_{\oplus}$, and an angular separation in arcsec, $\theta$, can be expressed as

\begin{equation}
s \simeq  2900 \left({\theta \over 0.1}\right) \left({\Delta_{\oplus} \over 40 }\right) {\rm \ km} \,,
\label{eqn:sep}
\end{equation}

\noindent or nearly 60 primary radii for a 100 km diameter TNO. This leaves a significant amount of separation phase space where a sizable population of stable binaries could exist undiscovered.

At the extreme of small separation are bilobed objects and contact binaries.  {\it Sheppard and Jewitt}, (2004) proposed that the TNO 2001 QG$_{298}$ may be a very close, possibly contact binary based on its large amplitude lightcurve (1.14 mag) and its relatively slow rotation (13.77 hr).  The rotation is too slow and the amplitude too large to explain the lightcurve as a fluid deformation into a Jacobi ellipsoid as is postulated for some other TNOs ({\it Sheppard and Jewitt}, 2002; {\it Jewitt and Sheppard}, 2002).   That leaves only albedo variation, strength-dominated shape, or a close binary to explain the lightcurve of 2001 QG$_{298}$.  {\it Sheppard and Jewitt}, (2004) conclude that a close binary is the most likely explanation for the observed lightcurve.  They further extrapolate from this single object to suggest that as many as 10--15\% of TNOs could be close binaries.  While unproven, a 10--15\% fraction of TNOs as contact binaries is not incompatible with the frequency of wider binaries and the trend of frequency as a function of separation.

The natural dimension for scaling the separations of TNBs is the Hill radius, $R_{\rm H}$, given by the equation

\begin{equation}
R_{\rm H} \cong a_{\odot} (M_1/3M_{\odot})^{1/3} \,,
\label{eqn:Hill}
\end{equation}

\noindent where $M_1$ is the mass of the largest component of the binary, $M_{\odot}$ is the mass of the Sun, and $a_{\odot}$ is the semimajor axis of the primary's heliocentric orbit.  Satellite orbits well inside the primary's Hill sphere are generally stable with respect to external perturbations.  However, the instantaneous Hill radius, calculated by replacing $a_{\odot}$ above with the instantaneous heliocentric distance, can deviate significantly from the average Hill radius for TNBs with large heliocentric eccentricities such as (65489) Ceto/Phorcys.  This result must be taken into account when considering, for example, the impact of gravitational perturbations from major planets.  The Hill radius scales with the radius of the primary and, thus, the large size range of the primary objects in Table 2 implies a similarly large range in Hill radii measured in absolute units.  In dimensionless units, a typical Hill radius for a TNB is on the order of 7000 times the radius of the primary.  Objects with measured orbits have semimajor axes that range from  $\sim$0.1\% to $\sim$8\% of the Hill radius, ({\it Noll et al.}, 2004b; {\it Kern}, 2005) well within the stable portion of the Hill sphere.  Whether this represents an intrinsic outcome of formation ({\it Astakhov et al.}, 2005), or is a signature of a surviving remnant population where more weakly bound systems have been disrupted while more tightly bound systems have become even tighter in the wake of encounters with third bodies ({\it Petit and Mousis}, 2004), or some combination of the two, remains to be determined.

\bigskip

\bigskip
\noindent
\textbf{ 3.4 Mass, Albebo and Density}
\bigskip

A particularly valuable piece of information that can be derived from the mutual orbit of a binary system is the total mass, $M_{\rm sys}$, of the system, according to the equation

\begin{equation}
M_{\rm sys} = \frac{4\pi^2 a^3}{GT^2} \,,
\label{eqn:msys}
\end{equation}

\noindent where $a$ is the semimajor axis, $G$ is the gravitational constant, and $T$ is the orbital period.  Knowledge of $a$ tends to be limited by the spatial resolution of the telescope, while knowledge of $T$ is limited by the timespan over which observations are carried out (modulo the binary orbit period).  It is possible to extend the timespan of observations, whereas the spatial resolution cannot generally be improved.  Thus, typically, $T$ is determined to much higher fractional precision than $a$, and the uncertainty in $M_{\rm sys}$ is dominated by the uncertainty in $a$.  $M_{\rm sys}$ can often be calculated before all 7 elements of the binary orbit are fully determined, since the elements $T$, $a$, and $e$ are relatively insensitive to the ambiguity between orbits mirrored through the instantaneous sky plane (see \S {3.1}).

For a system with a known mass, it is possible to derive other parameters which offer potentially valuable compositional constraints.  For instance, if one assumes a bulk density $\rho$, the bulk volume of the system $V_{\rm sys}$ can be computed as $V_{\rm sys}$ = $M_{\rm sys}$/$\rho$.  How the volume and mass is partitioned between the two components remains unknown.  Assuming the components share the same albedo, the individual radii of the primary and secondary can be obtained from

\begin{equation}
R_1 = \left( \frac{3V_{\rm sys}}{4\pi (1 - 10^{{-0.6\Delta_{\rm mag}}} ) } \right)^{1/3}
\label{eqn:rad1}
\end{equation}

and equation (\ref{eqn:relsize}) simplifies to give $R_2 = R_1 10^{-0.2\Delta_{\rm mag}}$.

An effective radius $R_{\rm eff}$, equal to the radius of a sphere having the same total surface area as the binary system can be computed as

\begin{equation}
R_{\rm eff} = \sqrt{ R_1^2 + R_2^2} \,.
\label{eqn:reff}
\end{equation}

Combining $R_{\rm eff}$ and the absolute magnitude of the system $H_{\lambda}$, one obtains the geometric albedo $p_{\lambda}$  

\begin{equation}
p_{\lambda} = \left( \frac{C_{\lambda}}{R_{\rm eff}} \right)^2 10^{-0.4 H_{\lambda}}
\label{eqn:albedo}
\end{equation}

\noindent  where $C_{\lambda}$ is a wavelength-dependent constant ({\it Bowell}, 1989; {\it Harris}, 1998).  For observations in the V band $C_V = {664.5\ {\rm km}}$.  This approach has been used to estimate albedos for a number of TNBs by assuming their bulk densities must lie within a plausible range, typically taken to be 0.5 to 2 g cm$^{-3}$ ({\it e.g., Noll et al.}, 2004a, 2004b; {\it Grundy et al.}, 2005; {\it Margot et al.}, 2005).  These efforts demonstrate that the TNBs have very diverse albedos, but those albedos are not clearly correlated with size, color, or dynamical class.  The calculation could also be turned around such that an assumed range of albedos leads to a range of densities.

When the sizes of the components of a binary system can be obtained from an independent observation, that information can be combined with the system mass to obtain the bulk density, providing a fundamental constraint on bulk composition and interior structure.  Sizes of TNOs are extremely difficult to obtain, although a variety of methods can be used, ranging from direct observation ({\it e.g., Brown et al.}, 2004) to mutual events and stellar occultations ({\it e.g., Gulbis et al.}, 2006).  For rotationally deformed bodies it is possible to constrain the density directly from the observed lightcurve assuming the object is able to respond as a ``fluid'' rubble pile ({\it Jewitt and Sheppard}, 2004; {\it Takahashi and Ip}, 2004; chapter by {\it Sheppard et al.}).  Spitzer Space Telescope observations of thermal emission have recently led to a number of TNO size estimates ({\it e.g., Cruikshank et al.}, 2006; chapter by {\it Stansberry et al.}).  Unfortunately, many of the known binaries are too small and distant to be detected at thermal infrared wavelengths by Spitzer or directly resolved by HST.  

Systems with density estimates include three large TNBs: Pluto and Charon, with $\rho = 2.0\pm$0.06 and $\rho = 1.65\pm$0.06 g cm$^{-3}$, respectively ({\it Buie et al.}, 2006), 2003 EL$_{61}$ with $\rho =  3.0\pm$0.4 g cm$^{-3}$ ({\it Rabinowitz et al.}, 2006),  and Eris with $\rho = 2.26\pm$0.25 g cm$^{-3}$ ({\it Brown}, 2006).  The relatively high densities of the large TNOs are indicative of substantial amounts of rocky and\slash or carbonaceous material in the interiors of these objects, quite unlike their ice-dominated surface compositions.

Four smaller TNBs have recently had their densities determined from Spitzer radiometric sizes.  These are (26308) 1998 SM$_{165}$ with $\rho = 0.70\pm {0.32 \atop 0.21}$ g~cm$^{-3}$ ({\it Spencer et al.} 2006), (47171) 1999 TC$_{36}$ with $\rho = 0.5\pm {0.3 \atop 0.2}$ g~cm$^{-3}$ ({\it Stansberry et al.}, 2006), (65489) Ceto\slash Phorcys with $\rho = 1.38\pm {0.65 \atop 0.32}$ g~cm$^{-3}$ ({\it Grundy et al.}, 2007), and (42355) Typhon\slash Echidna with $\rho =  0.47\pm {0.18 \atop 0.10}$ g~cm$^{-3}$ ({\it Grundy et al.}\ in preparation).  {\it Takahashi and Ip} (2004) estimate a density of $\rho <$0.7 g cm$^{-3}$ for 2001 QG$_{298}$.  The very low densities of four of these five require little or no rock in their interiors, and even for pure H$_2$O ice compositions, call for considerable void space.  The higher density of Ceto/Phorcys is consistent with a mixture of ice and rock. It is clear from these results that considerable diversity exists among densities of TNOs, but it is not yet known whether densities correlate with externally observable characteristics such as color, lightcurve amplitude, or dynamical class.

\bigskip

\bigskip
\noindent
\textbf{ 3.5 Eccentricity Distribution and Tidal Evolution}
\bigskip

The eccentricities of binary orbits known to date span the range from values near zero to a high of 0.8 (Table 2), with perhaps a clustering in the range 0.3--0.5.  With the usual caveats about small number statistics (N$\sim$10), no obvious correlation between eccentricity and semimajor axis is present in the data obtained to date.  

Keplerian two-body motion assumes point masses orbiting one another.  The finite size of real binary components allows differential gravitational acceleration between nearer and more distant parts of each body to stretch them along the line connecting them.  If their mutual orbit is eccentric, this tidal stretching varies between apoapsis and periapsis, leading to periodic flexing.  The response of a body to such flexing is characterized by the parameter $Q$, which is a complex function of interior structure and composition ({\it Goldreich and Soter}, 1966; {\it Farinella et al.}, 1979).  In a body with low $Q$, tidal flexing creates more frictional heating, which dissipates energy.  A body with higher $Q$ can flex with less energy dissipation.  For a body having a rotation state different from its orbital rotation $Q$ is related to the angular lag, $\delta$, of the tidal bulge behind the line of centers: $Q^{-1}$ = tan(2$\delta$).  As with orbital eccentricity, this situation produces time-variable flexing, and thus frictional heating and dissipation of energy.  Typical values of $Q$ for rocky planets and icy satellites are in the 10 to 500 range ({\it Goldreich and Soter}, 1966; {\it Farinella et al.}, 1979; {\it Dobrovolskis et al.}, 1997).   

The energy dissipated by tidal flexing comes from orbital and/or rotational energy, leading to changes in orbital parameters and rotational states over time.  Tides raised on the primary tend to excite eccentricity, while tides raised on the secondary result in damping. The general trend is toward circular orbits, with both objects' spin vectors aligned and spinning at the same rate as their orbital motion.  The timescale for circularization of the orbit is given by

\begin{equation} 
\tau_{\rm circ} = \frac{4Q_2M_2}{63M_1} \sqrt{ \frac{a^3}{G(M_1 + M_2)} } 
\left( \frac{a}{R_2} \right)^5 
\label{eqn:simplified}
\end{equation}

\noindent where $a$ is the orbital semimajor axis, $M_1$ is the mass of the primary, and $M_2$, $R_2$, and $Q_2$ are the mass, radius, and $Q$ of the secondary ({\it e.g., Goldreich and Soter}, 1966).  It is important to recognize that this formula (a) assumes the secondary to have zero rigidity, and (b) assumes that the eccentricity evolution due to tides raised on the primary is ignorable. Neither
of these assumptions may be justified, especially in the case of near-equal sized binaries (for a thorough discussion, see {\it Goldreich and Soter}, 1966).
The timescale (\ref{eqn:simplified}) is sensitive to the ratio of the semimajor axis to the size of the secondary.  Larger and/or closer secondaries are likely to have their orbits circularized much faster than more widely-separated systems.  For example, the close binary (65489) Ceto/Phorcys has $a$ = 1,840 km,  $M_1$ = 3.7$\times 10^{18}$ kg, $M_2$ = 1.7$\times 10^{18}$ kg, and $R_2$ = 67 km ({\it Grundy et al.}, 2007).  For $Q$ = 100 (a generic value for solid bodies), its orbit should circularize on a relatively short timescale of the order of $\sim$10$^5$ years.  A more widely separated example, (26308) 1998 SM$_{165}$ has $a$ = 11,300 km, $M_1$ = 6.5$\times 10^{18}$ kg, $M_2$ = 2.4$\times 10^{17}$ kg,  and $R_2$ = 48 km ({\it Spencer et al.}, 2006), leading to a much longer circularization timescale of the order of $\sim$10$^{10}$ years, consistent with the observation that it still has a moderate orbital eccentricity of 0.47 ({\it Margot}, 2004).

Tidal effects can also synchronize the spin rate of the secondary to its orbital period (as in the case of the Earth/Moon system) and, on a longer timescale, synchronize the primary's spin rate as well (as for Pluto/Charon).  The timescale for spin locking the primary (slowing its spin to match the mutual orbital period) is given by

\begin{equation} 
\tau_{{\rm despin},1} = \frac{Q_1 R_1^3 \omega_1}{GM_1} \left( 
\frac{M_1}{M_2} \right)^2 \left( \frac{a}{R_1} \right)^6 
\label{eqn:despin1}
\end{equation}

\noindent where $\omega _1$ is the primary's initial angular rotation rate and $R_1$ is its radius ({\it Goldreich and Soter}, 1966).  The initial angular rotation rate is not known, but an upper limit is the break-up rotation rate, which is within 50 percent of 3.3 hours for the 0.5 to 2 g cm$^{-3}$ range of densities discussed. For (65489) Ceto/Phorcys and (26308) 1998 SM$_{165}$ with primary radii ($R_1$) of 86 km ({\it Grundy et al.}, 2007) and 147 km ({\it Spencer et al.}, 2006), we find the spinlock timescale to be  $\sim$10$^4$ and $\sim$10$^5$ years respectively, slightly faster than the circularization timescale. The timescale for de-spinning the secondary, $\tau_{\rm despin,2}$, is given by swapping subscripts $1$ and $2$ in (\ref{eqn:despin1}).

The general case for both tidal circularization and tidal despinning in binaries is complex ({\it e.g. Murray and Dermott}, 1999).  Binaries in the transneptunian population include many where the secondary is of comparable size to the primary.  For these systems, it is not safe to make the common assumption that the tide raised by the secondary on the primary is ignorable.  Furthermore, there are many systems where the binary orbit has moderate to high eccentricity invalidating the simplifying assumptions possible for nearly circular orbits.  A full treatment of the tidal dynamics for the kinds of systems we find in the transneptunian population would make an interesting addition to the literature.  In the meantime, observation is likely to lead the way in understanding the dynamics of these systems.

\bigskip

\bigskip
\noindent
\textbf{ 3.6 Colors and Lightcurves}
\bigskip

The spectral properties of TNOs and their temporal variation are fundamental probes of the surfaces of these objects (see chapters 7-11, this volume).   The colors of TNOs have long been known to be highly variable ({\it Jewitt and Luu}, 1998) and some correlations of color with other physical or dynamical properties have been claimed (e.g. {\it Peixinho et al.}, 2004).  A natural question is whether this variability can be used to constrain either the origin of TNBs, the origin of color diversity, or both.  For example, one can ask whether TNB components are similarly or differently colored.  Because TNBs are thought to be primordial, differences in color could be either due to mixing of different-composition populations in the protoplanetary disk before the bound systems were formed or different collisional and evolutionary histories of components after they are bound. 

A handful of TNBs have reported single-epoch resolved color measurements.  Some of these objects are solar-colored [2000 CF$_{105}$, (58534) Logos/Zoe, (47171) 1999 TC$_{36}$, (66652) 1999 RZ$_{253}$, (88611) 2001QT$_{297}$], while others are more gray [2001QC$_{298}$, (65489) Ceto/Phorcys].  However, so far, the components have colors that are consistent with each other within the uncertainties of the measurements, 0.1--0.3 mags ({\it Margot}, 2005; {\it Noll et al.}, 2004a,b; {\it Osip et al}, 2003; {\it Grundy et al.}, 2007).  This similarity implies that the components are composed of similar material, at least on the surface.  It also suggests that the assumption of equal albedo and density usually made for binaries may have some basis in fact. 

Spectra are even better composition diagnostics than color measurements.  Separate spectra of binary components are currently available only for the Pluto/Charon pair ({\it Buie et al.}, 1987; {\it Fink and DiSanti}, 1988; {\it Buie and Grundy}, 2000) and for 2003 EL$_{61}$ ({\it Barkume et al.}, 2006).  Pluto and Charon have well known spectral differences that may be primarily related to the size threshold for retaining the very volatile CH$_4$ and N$_2$ ices found on Pluto but not on Charon.  2003 EL$_{61}$ and its larger satellite, by contrast, both have spectra that are dominated by water ice.

Lightcurves are diagnostic of both compositional variation on surfaces and of non-spherical shapes.  They also give rotation rates.  In binary systems, the rotation state of the components is subject to tidal evolution (see \S 3.5).  Both unresolved and resolved lightcurves can be useful for addressing these issues.

Unresolved lightcurves of short duration for a number of TNBs have been obtained, sometimes with incomplete or inconsistent results.  Lightcurves of (47171) 1999 TC$_{36}$ and (42355) Typhon/Echidna showed variations on the order of 0.10--0.15 mags, but no period was determinable from the data ({\it Ortiz et al.}, 2003).  Similarly, observations of (66652) 1999 RZ$_{253}$ and 2001 QC$_{298}$ showed small amplitude, but non-systematic variation over a 4--6 hour duration ({\it Kern}, 2005).  {\it Romanishin et al.} (2001) reported a lightcuve for the unresolved binary (26308) 1998 SM$_{165}$, obtained from the 1.8m Vatican Advanced Technology Telescope in 1999 and 2000, with a moderate amplitude of 0.56 mags.  The period was determined to be either 3.983 hours (single-peaked, caused by an albedo spot) or 7.966 hours (double-peaked, caused by nonspherical shape).  The single-peaked period is near the break-up period (~3.3 hours) for a solid ice body.   Because the unresolved lightcurve of (26308) 1998 SM$_{165}$ did not show any color variation with time {\it Romanishin et al.} argued for the longer, double-peaked period as the most likely.  Lightcurve measurements of the same binary made at Lowell Observatory in 2006 found a slightly longer period  of 8.40$\pm$0.05 hours ({\it Spencer et al.}, 2006).   

Resolved ground based lightcurves of binaries are challenging and can only be obtained under excellent conditions at a few facilities, and only for objects with the widest separations. Discovery observations of the binary (88611) 2001 QT$_{297}$ at Las Campanas Observatory with Magellan revealed brightness changes in the secondary component of 0.3 mag in 30 minutes ({\it Osip et al.}, 2003).  Follow-up observations revealed the secondary to have a single-peaked period of 5--7 hours while the magnitude of the primary remained constant. Additional resolved color lightcurve measurements found the two surfaces to share similar colors throughout the rotation indicating homogeneous, similar surfaces ({\it Kern}, 2005).  Similar observations showed that both components of 2003 QY$_{90}$ to be variable.  The primary and secondary components were observed to change by 0.34$\pm$0.12 and 0.9$\pm$0.36 mags, respectively, over six hours of observation ({\it Kern and Elliot}, 2006).  The large amplitude of the secondary component sometimes results in the secondary being brighter than the primary.  Both components of the wide binary 2005 EO$_{304}$ are variable with variations on the order of 0.3 mags over a period of 4 hours ({\it Kern}, 2005).

Space based observations from HST resolve the components of binaries and can constrain the variability of components in these systems.  The best studied system, by far, is the Pluto/Charon binary where detailed lightcurve measurements have been made ({\it Buie et al.}, 1997).  TNBs that have had their orbits measured by HST have multiple-epoch photometric measurements, although frequently the temporal sampling is poor.  (58534) Logos/Zoe shows variability in the primary of at least $\sim$ 0.8 mag, making it challenging at times to distinguish the primary from the secondary ({\it Noll et al.}, 2004).  However, with only a few widely spaced samples, this remains, for the moment, only an intriguing suggestion of a lightcurve.  Three objects, (47171) 1999 TC$_{36}$, 2001 QC$_{298}$ and (65489) Ceto/Phorcys have virtually no variation in flux, implying they may be relatively spherical, homogeneous, and/or pole-on  (although we note the contradictory ground-based observations of (47171) 1999 TC$_{36}$).  Once again, the sampling density is far to small to allow anything more than informed speculation.

\bigskip

\bigskip
\noindent
\textbf{ 3.7 Orbit Plane and Mutual Events}
\bigskip

The tremendous scientific benefit that can derive from mutual occultations or eclipses between a poorly-resolved object and its satellite was abundantly illustrated by the series of mutual events between Pluto and Charon during the 1980s ({\it Binzel and Hubbard}, 1997).  As discussed before, these events enabled measurement of the sizes and albedos of both objects, of their distinct surface compositions, and even of albedo patterns on their surfaces.

For observable mutual events to happen either the observer or the Sun (or both) must be temporarily aligned with a TNB's orbit plane.  An ``occultation-type'' event occurs when one component of the TNB passes in front of, and fully or partially occults, the other component from the observer's point of view.  An ``eclipse-type'' event takes place when the TNB components are aligned with the Sun and the shadow of one falls on the other.  Because the Sun and Earth have nearly equal lines of sight to TNBs, most mutual events  observable from the Earth are combinations of occultation-type and eclipse-type events.  

The larger the objects are compared with their separation, the farther the orbit plane can deviate from either of these two types of alignments and still produce an observable mutual event.  The criteria for both types of events can be expressed as $R_1 + R_2 > s\ {\rm sin}(\phi)$ where $R_1$ and $R_2$ are previously defined and, $s$ is their separation during a conjunction (equal to the semimajor axis, for the case of a circular orbit), and $\phi$ is the angle between the observer or the Sun and the plane of the binary orbit.  During any conjunction when either criterion is satisfied, a mutual event can be observed.  The period during which the orbit plane is aligned closely enough to the Sun's or to the Earth's line of sight to satisfy the criteria for events can be thought of as a mutual event season.

Each orbit of a transneptunian secondary brings a superior conjunction (when the primary is closer to the observer) and an inferior conjunction (when the secondary is closer), so shorter orbital periods lead to more frequent conjunctions and associated opportunities for mutual events.  The most recent mutual event season of Pluto and Charon lasted from 1985 through 1990, and, since their mutual orbit period is only 6.4 days, there were hundreds of observable events during that season.  For more widely separated, smaller pairs, with longer orbital periods, the mutual event seasons may be shorter and conjunctions may be less frequent, leading to far fewer observable events, or even none at all.  For example, (26308) 1998 SM$_{165}$ has a reasonably well-determined orbit with a period of 130 days, a semimajor axis of 11,170 km , and an eccentricity of 0.47 ({\it Margot}, 2004).  From Spitzer thermal observations, the diameters of the primary and secondary bodies are estimated to be 294 and 96~km, respectively ({\it Spencer et al.}, 2006).  Ignoring uncertainties in the current orbital elements, the mutual event season will last from 2020 through 2026, with 12 mutual events being observable at solar elongations of 90 degrees or more.  Of these 12 events, 2 are purely occultation-type events and one is purely an eclipse-type event.  The rest involve combinations of both occultation and eclipse.

\bigskip

\section{\textbf{BINARY FORMATION}}

When the Pluto/Charon binary was the only example of a true binary in the solar system (true binary $\equiv$ two objects orbiting a barycenter located outside either body) it could be discounted as just another of the peculiarities associated with this yet-to-be-dwarf planet.  The discovery of numerous similar systems among TNOs, however, has changed this calculus.  Any successful model for producing TNBs cannot rely on low-probability events, but must instead employ processes that were commonplace in the portion of the preplanetary nebula where these objects were formed.  Formation models must also account for the observed properties of TNBs including the prevalence of similar-size binaries, the range of orbital eccentricities, and the steeply increasing fraction of binaries at small angular separations.  Survival of binaries, once they are formed, is another important factor that must be considered when, for instance, comparing the fraction of binaries found in different dynamical populations or their distribution as a fraction of Hill radius.  

Several possible modes for the formation of solar system binaries have been discussed in the literature including fission, dynamical capture, and collision ({\it c.f.~reviews by Richardson and Walsh}, 2006; {\it Dobrovolskis et al.}, 1997).  For TNOs, capture and/or collision models have been the most thoroughly investigated.  Fission is unlikely to be important for objects as large as the currently known TNBs.  Other possible mechanisms for producing binaries, {\it e.g.} volatile-driven splitting, as is observed in comets, have not been explored.  Interestingly, both capture and collisional formation models share the requirement that the number of objects in the primordial Kuiper Belt (at least the small ones) be at least a couple of orders of magnitude higher than currently found in transneptunian space.  It follows that all of the TNBs observed today are primordial.

\bigskip

\bigskip
\noindent
\textbf{ 4.1 Capture}
\bigskip

Capture models rely, in one form or another, on three-body interactions to remove angular momentum and produce a bound pair.  As we show in detail below in \S 4.5, they are also very sensitive to the assumed velocity distribution of planetesimals.  {\it Goldreich et al.} (2002, hereafter {\it G02}) described two variations of the three-body model, $L^3$, involving three discrete bodies, and $L^2s$, where the third body is replaced by a dynamical drag coefficient corresponding to a ``sea'' of weakly interacting smaller bodies.  In {\it G02's} analysis, the $L^2s$ channel was more efficient at forming binaries by roughly an order of magnitude.  

{\it Astakhov et al.}~(2005) extended the capture model by exploring how a weakly and temporarily bound ($a\sim R_{\rm H}$) pair of big bodies hardens when a third small body (``intruder'') passes within the Hill radius (\ref{eqn:Hill}), of the pair.  Their calculations assume the existence of transitory binaries that can complete up to $\sim$10 mutual orbits before the third body approaches.  They find that a binary hardens most effectively when the intruder mass is a few percent that of a big body (this result likely depends on their assumed approach velocities of up to $5 v_{\rm H}$ where  $v_{\rm H} \equiv \Omega_{\rm K} R_{\rm H}$ is the Hill velocity, with $\Omega_{\rm K} \simeq 2\pi/(200\yr)$ denoting the local Kepler frequency).

The two capture formation channels $L^3$ and $L^2s$ require that binaries form---and formation times increase with decreasing separation $a$---before $v > v_{\rm H}$.  Given the preponderance of classical binaries ({\it Stephens and Noll}, 2006; Fig.), we can speculate that the primordial classical belt may have enjoyed dynamically cold (sub-Hill or marginally Hill) conditions for a longer duration than the primordial scattered population.  

Dynamical capture is the only viable formation scenario for many TNBs with high angular momentum (\S 4.4) and is a possible formation scenario for most, if not all, known TNBs.  Given the apparent importance of capture models, we explore them in detail in \S 4.5 with a specific focus on the case of binaries with similar mass components.

\bigskip
\bigskip
\noindent
\textbf{ 4.2 Collision}
\bigskip

Collisional models were proposed for the Pluto/Charon binary early on based on the angular momentum of the system ({\it McKinnon}, 1984, 1989) and one such model has recently been shown to be numerically feasible ({\it Canup}, 2005).  However, as we note below, angular momentum arguments alone are not sufficient to prove an impact origin.  Nix and Hydra, the outermost satellites of Pluto, plausibly resulted from the same impact that generated Charon ({\it Stern et al.}, 2006).  {\it Ward and Canup} (2006) propose that collisional debris within the exterior 4:1 and 6:1 resonances of Charon---resonances stabilized by Charon's initially large eccentricity---accumulated to form Nix and Hydra.  According to their scenario, as Charon's orbit tidally expanded, the small satellites would have been forced outwards to their current locations in resonant lockstep with Charon. Tidal circularization of Charon's orbit would have eventually released Nix and Hydra from resonance.  In this scenario, the nearly circular orbits of Nix and Hydra result from their coalescence from a dissipative, nearly circular disk; their eccentricities are not altered by resonant migration because the resonances involved are of the co-rotation type.

The small satellites of Pluto-sized TNOs 2003 EL$_{61}$ and (136199) Eris, characterized by satellite-to-primary mass ratios of $\sim$1\%, might also have formed by impacts ({\it Stern}, 2002).  Some collisional simulations ({\it Durda et al.}, 2004; {\it Canup}, 2005) reproduce such low mass ratios.  Tidal expansion of satellite orbits can explain, to within factors of a few, the current semimajor axes of the companions of 2003 EL$_{61}$ ({\it Brown et al.} 2005a).  An unresolved issue is the origin of the small, but significant, orbital eccentricity, $0.05 \pm 0.003$, for the outermost satellite of 2003 EL$_{61}$.  Tides should have reduced the eccentricity to values much smaller.  Also the mutual orbital inclination of the satellites of 2003 EL$_{61}$, which might be as large as $39^{\circ}$ ({\it Brown et al.} 2006a), has yet to be explained.  Formation by collisionless capture along the lines of {\it G02}, though not ruled out, remains poorly explored for unequal mass components ({\it Brown et al.} 2005a).

To occur with reasonable frequency, collisions between Pluto-sized ($R\sim 1000\km$) TNOs must be gravitationally focused.  Transneptunian space may have been populated by a few dozen such objects ({\it Kenyon and Luu} 1999).  If their relative velocities were less than the Hill velocity, $v_{\rm H}$, then the collision timescale would be $\sim$$6\Myr$ (\ref{eqn:sub}).  Otherwise the collision time exceeds $\sim$500 Myr.  Like collisionless capture (see \S {4.1}), binary formation by giant impacts must have taken place while the disk was dynamically cold.  

\bigskip
\bigskip

\noindent
\textbf{ 4.3 Hybrids}
\bigskip

Hybrid collision/capture models are possible as well; two variants on this theme have been proposed.  

{\it Weidenschilling} (2002) considered a model in which a third big body collides with one member of the scattering pair.  Since physical collisions have smaller cross-sections than gravitational interactions, this mechanism requires $\sim$$10^2$ more big ($R \sim 100\km$) bodies than are currently observed to operate at the same rate as $L^3$.  It also predicts an unobserved prevalence of widely separated binaries.  

{\it Funato et al.} (2004) proposed that observed binaries form by the exchange reaction $Ls+L\rightarrow L^2 +s$ wherein a small body of mass $M_{\rm sm}$, originally orbiting a big body of mass $M_{\rm big}$, is ejected by a second big body.  In the majority of ejections, the small body's energy increases by its orbital binding energy $\sim$$M_{\rm sm}v_{\rm esc}^2/2$, leaving the big bodies bound with separation $a \sim (M_{\rm big}/M_{\rm sm})R$.   As formulated, this model predicts a prevalence of very high eccentricity binaries that is not observed.   The rate-limiting step for the exchange reaction model is the formation of the pre-existing ($Ls$) binary, which requires two big bodies to collide and fragment  (\ref{eqn:sub}).

\bigskip

\bigskip
\noindent
\textbf{ 4.4 Angular Momentum}
\bigskip

It is possible, in some cases, to distinguish between capture and collision based on the angular momentum of the binary.  Early theoretical arguments in favor of a collisional origin for Pluto/Charon were based on the fact that the total angular momentum of the system exceeds breakup for a single, reconstituted object ({\it e.g.~McKinnon}, 1989).  However, it can also be shown that to have formed from a fragmentary collision, binary components cannot have too much angular momentum.  It is conventional to express angular momentum as $J/J'$, where the combined orbital and spin angular momentum of the binary $J$ is normalized by $J' = \sqrt{GM_{\rm tot}^3R_{\rm eff}}$, where $G$ is the gravitational constant, $M_{\rm tot}$ is the total system mass, and $R_{\rm eff}$ is the radius of an equivalent spherical object containing the total system mass. {\it Canup} (2005) found that binary systems produced by single collisions have $J/J' < 0.8$ (for an order-of-magnitude derivation of this result, see {\it Chiang et al.}, 2006).  For instance, the Earth/Moon system has $J/J' \simeq 0.1$ and the Pluto/Charon system has $J/J'\simeq 0.4$.  

In Table 2, we list  $J/J'$ for TNBs.  We use actual measurements where they exist, otherwise the calculation assumes that both binary components have spin periods of 8 hours and densities of 1 g cm$^{-3}$.  For about half of the TNBs documented in Table 2, $J/J'$ exceeds unity, so much angular momentum that formation via two-body collisions can be ruled out.

\bigskip

\bigskip
\noindent
\textbf{ 4.5 Detailed Capture Models}
\bigskip

Capture models are of particular importance for TNBs as the only class of models that can explain the existence of high angular momentum systems.  We review and expand, in detail, on capture models in this section. 

TNOs can become bound (``fuse'') by purely gravitational means while they are still dynamically cold.  Following {\it G02}, we consider how ``big'' TNOs, having sizes
$R_{\rm big} \sim 100$ km, fuse when immersed in a primordial sea of ``small'' bodies, each of size $R_{\rm sm}$.  We assume that the small bodies contain the bulk of the disk mass: the surface density of small bodies $\sigma \sim \sigma_{\rm MMSN}$, where $\sigma_{\rm MMSN} \sim$ 0.2 g cm$^{-2}$
is the surface density of solids in the minimum-mass solar nebula at a heliocentric distance of 30 AU.  
We assume that the surface density of big bodies was the same then as it is now: $\Sigma \sim 0.01 \sigma_{\rm MMSN}$. This last condition agrees with the output of numerical simulations of coagulation by {\it Kenyon and Luu} (1999).  The velocity dispersion of small bodies is $u > v_{\rm H}$.  For convenience we define $\alpha \equiv R_{\rm big}/R_{\rm H} \simeq 1.5 \times 10^{-4}$ and note that the surface escape velocity from a big body $v_{\rm esc} \simeq v_{\rm H} \alpha^{-1/2}$.  The velocity dispersion of big bodies is $v < u$.

Small bodies have their random velocities $u$ amplified by gravitational stirring by big bodies and damped by inelastic collisions with other small bodies.  Balancing stirring with damping sets $u$ ({\it G02}):

\begin{equation}
\frac{u}{v_{\rm H}} \sim \left( \frac{R_{\rm sm}}{R_{\rm big}} \frac{\Sigma}{\sigma} \right)^{1/4} \alpha^{-1/2} \sim 3 \left( \frac{R_{sm}}{20 \m} \right)^{1/4} \,.
\label{eqn:vsie}
\end{equation}

\noindent Big bodies have their random velocities $v$ amplified by gravitational stirring by other big bodies and damped by dynamical friction with small bodies. When $v > v_{\rm H}$, this balance yields
({\it Goldreich et al.}~2004, hereafter {\it G04}):

\begin{equation}
\frac{v}{u} \sim \left( \frac{\Sigma}{\sigma} \right)^{1/4} \sim \frac{1}{3} \,.
\label{eqn:ve}
\end{equation}

\noindent Combining (\ref{eqn:vsie}) with (\ref{eqn:ve}), we have

\begin{equation}
\frac{v}{v_{\rm H}} \sim \left( \frac{\Sigma}{\sigma} \right)^{1/2} \left( \frac{R_{\rm sm}}{R_{\rm big}} \right)^{1/4} \alpha^{-1/2} \sim 1 \left( \frac{R_{\rm sm}}{20\m} \right)^{1/4} \,,
\label{eqn:vvH}
\end{equation}

\noindent valid for $R_{\rm sm} > 20 \m$. If $R_{\rm sm} < 20 \m$, then $v < v_{\rm H}$, neither (\ref{eqn:ve}) nor (\ref{eqn:vvH}) holds, but (\ref{eqn:vsie}) still does.

By allowing for the possibility that $v > v_{\rm H}$, we depart from {\it G02}.  When $v > v_{\rm H}$, inclinations and eccentricities of the big bodies' heliocentric orbits can be comparable ({\it G04}).  If, prior to fusing, big bodies have an isotropic velocity dispersion, then the resultant mutual binary orbits will be randomly inclined, in agreement with observation. Otherwise, if $v < v_{\rm H}$, big bodies collapse into a vertically thin disk ({\it G04}) and mutual orbit normals, unless subsequently torqued, will be parallel, contrary to observation.  Invoking $v > v_{\rm H}$ comes at a cost: Efficiencies for fusing decrease with increasing $v$.  To quantify this cost, we define a normalized velocity parameter $F$ as

\begin{equation}\label{eqn:F}
F =  
   \left\{\begin{array}{ll}
     1  &\mbox{if $v < v_{\rm H}$} \\  
    \rule{0ex}{5ex}v/v_{\rm H} &\mbox{if $v > v_{\rm H}$}   
  \end{array}\right.  
\end{equation}  

\noindent and derive how the rates of fusing depend on $F$.  How $F$ increased to its current value of $\sim$$10^3$---i.e., how the Kuiper belt was dynamically heated---remains contested ({\it Chiang et al.}, 2006; {\it Levison et al.}, 2006). 

Both the $L^3$ and $L^2s$ scenarios described by {\it G02} begin when one big body ($L$) enters a second big body's ($L$) sphere of influence.  This sphere has radius $R_{\rm I} \sim R_{\rm H} / F^2$.  Per big body, the entry rate is

\begin{equation}
\dot{N}_{\rm I} 
\sim \frac{\Sigma \Omega_{\rm K}}{\rho R_{\rm big}} \alpha^{-2} F^{-4}\,.
\label{eqn:enter}
\end{equation}

\noindent If no other body participates in the interaction, the two big bodies pass through their spheres of influence in a time $t_{\rm enc} \sim R_{\rm I}/v \sim \Omega_{\rm K}^{-1} F^{-3}$ (assuming they do not collide).  The two bodies fuse if they transfer enough energy to other participants during the encounter.  In $L^3$, transfer is to a third big body: $L+L+L\rightarrow L^2+L$.  To just bind the original pair, the third body must come within $R_{\rm I}$ of the pair.  The probability for this to happen in time $t_{\rm enc}$ is $P_{L^3} \sim \dot{N}_{\rm I} t_{\rm enc}$.  If the third body succeeds in approaching this close, the probability that two bodies fuse is estimated to be on the order of unity (this probability has yet to be precisely computed).  Therefore the timescale for a given big body to fuse to another by $L^3$ is

\begin{equation}
t_{{\rm fuse}, L^3} \sim \frac{1}{\dot{N}_{\rm I} P_{L^3}} \sim \left( \frac{\rho R_{\rm big}}{\Sigma} \right)^2 \frac{\alpha^4}{\Omega_{\rm K}} F^{11} \sim 2 F^{11} \Myr \,.
\label{eqn:fuse_L3}
\end{equation}

\noindent
The extreme sensitivity to $F$ in equation (\ref{eqn:fuse_L3}) implies that no binaries form by $L^3$ once $v$ exceeds $v_{\rm H}$.

In $L^2s$, energy transfer is to small bodies by dynamical friction: $L+L+s^\infty\rightarrow L^2+s^\infty$.  In time $t_{\rm enc}$, the pair of big bodies undergoing an encounter lose a fraction $(\sigma \Omega_{\rm K}/ \rho R_{\rm big}) (v_{\rm esc}/u)^4 t_{\rm enc}$ of their energy, under the assumption $v_{\rm esc} > u > v_{\rm H}$ ({\it G04}).  This fraction is on the order of the probability $P_{L^2s}$ that they fuse, whence

\begin{eqnarray}
t_{{\rm fuse},L^2s} & \sim & \frac{1}{\dot{N}_{\rm I} P_{L^2s}} \sim \left({\rho R_{\rm big} \over\sigma}\right)^{2} {R_{\rm sm}\over R_{\rm big}} \frac{\alpha^{2}}{\Omega_{\rm K}} F^{7} \nonumber \\
 & \sim & 1 \left( \frac{R_{\rm sm}}{20\m} \right) F^7 \Myr \,,
\label{eqn:fuse_L2S}
\end{eqnarray}

\noindent where we have used (\ref{eqn:vsie}).  The steep dependence on $F$ implies that formation by $L^2s$ remains viable only for $v/v_{\rm H}$ less than a few.  Together equations (\ref{eqn:fuse_L3}) and (\ref{eqn:fuse_L2S}) imply that explaining random binary inclinations by appealing to $v \gtrsim v_{\rm H}$ spawns a fine-tuning problem: Why should $v \approx v_{\rm H}$ during binary formation?

Having formed with semimajor axis $a\sim R_{\rm I}\sim (7000 / F^2)R_{\rm big}$, the mutual orbit shrinks by further energy transfer.  If $L^3$ is the more        efficient formation process, passing big bodies predominantly harden the binary; if $L^2s$ is more efficient, dynamical friction dominates hardening.  The probability $P$ per orbit that $a$ shrinks from $\sim$$R_{\rm I}$ to $\sim$$R_{\rm I}/2$ is of order either $P_{L^3}$ or $P_{L^2s}$.  We equate the formation rate of binaries, $N_{\rm all}/t_{\rm fuse}$, with the shrinkage rate, $\Omega_{\rm K} P N_{{\rm bin}}|_{x\sim R_{\rm I}}$, to conclude that the steady-state fraction of TNOs that are binaries with separation $R_{\rm I}$ is

\begin{equation}
f_{\rm bin} (a \sim R_{\rm I}) \equiv \frac{\left. N_{\rm bin}\right|_{a\sim R_{\rm I}}}{N_{\rm all}} \sim \frac{\Sigma}{\rho R_{\rm big}} \alpha^{-2} F^{-4} \sim 0.4 F^{-4}\% \,.
\end{equation}

\noindent As $a$ decreases below $R_{\rm I}$, shrinkage slows. Therefore $f_{\rm bin}$ increases with decreasing $a$. Scaling relations can be derived by arguments similar to those above. If $L^2s$ dominates, $f_{\rm bin} \propto a^0$ for $a < R_{\rm H} (v_{\rm  H}/u)^2$ and $f_{\rm bin} \propto a^{-1}$ for $a > R_{\rm H} (v_{\rm  H}/u)^2$ ({\it G02}).  If $L^3$ dominates, $f_{\rm bin} \propto a^{-1/2}$ when $v < v_{\rm H}$ and $f_{\rm bin} \propto a^{-1}$ when $v > v_{\rm  H}$.  For reference, resolved TNBs typically have $a\sim 100 R_{\rm big}$.  The candidate close binary reported by {\it Sheppard and Jewitt} (2004), and any similar objects, may be the hardened end-products of $L^3$ and $L^2s$ (though a collisional origin cannot be ruled out since for these binaries $J/J'$ might be less than unity).

These values for $f_{\rm bin}(a)$ characterize the primordial disk.  Physical collisions with small bodies over the age of the solar system, even in today's rarefied environment, can disrupt a binary.  For this reason, {\it Petit and Mousis} (2004) find that the widest and least massive binaries, having $a \gtrsim 400 R_{\rm big}$, may originally have been ten times more numerous than they are today.

Using the formalism we have developed for capture, it is also possible to develop expressions for the collision timescale.  As with capture, this rate of collisions is sensitive to the relative velocity of potential impactors, $F$.  This results from the dependence of the collisional rate on gravitational focusing.  For the two different velocity regimes, the timescales are given by

\begin{equation}\label{eqn:sub}  
t_{{\rm fuse,exchange}} \sim  
   \left\{\begin{array}{ll}
    \frac{\rho R}{\Sigma \Omega_k} \alpha^{3/2}  \sim 0.6 \Myr &\mbox{if $v < v_{\rm H}$} \\  
    \rule{0ex}{5ex} \frac{\rho R}{\Sigma \Omega_k} \alpha F^{2} \sim 50 F^{2} \Myr &\mbox{if $v > v_{\rm H}$.}
  \end{array}\right.  
\end{equation}

\bigskip

\section{\textbf{THE FUTURE OF BINARIES}}

Binaries offer unique advantages for the study of the Kuiper Belt and are likely to be among the most intensively studied TNOs.  Statistical studies will continue to refine the range of properties of binaries, their separations, their relative sizes, and their frequency as a function of dynamical class, size, and other physical variables.  The frequency of multiple systems will be constrained.  As binary orbits are measured, orbital parameters such as semimajor axis, eccentricity, and orbit plane orientation will also become the subject of statistical investigations.  System masses derived from orbits will fuel vigorous studies of physical properties of objects, both singly and in ensemble.  Albedo and density are constrained by the measurement of system mass alone, and can be separated with the addition of thermal infrared measurements.  The internal structure of TNOs can be inferred from densities.  The extremely low densities of a few objects measured so far requires structural models with a high fraction of internal void space.  The study of lightcurves of binaries is at a very early stage and can be expected to shed light on the shapes, pole orientations, and tidal evolution of binaries.  This, in turn, may yield additional information on the internal structure of TNOs by constraining the possible values of $Q$.

There are probably more than 100 binary systems that could be discovered in the currently-known transneptunian population of 600+ objects with well-established heliocentric orbits.  As the TNO population expands, the number of binaries can be expected to expand with it.  Most of these discoveries will be made with HST or other instruments with equivalent capabilities.  There are also yet-to-be explored areas of interest where theory and modeling can be expected to make significant progress.  As progress is made in understanding how binaries formed in the Sun's protoplanetary disk, these principles can be extended to other circumstellar disks that are now found in abundance.  If binary protoplanets are common, as seems to be the case for the solar system, we may even expect to find binary planets as we explore extrasolar planetary systems.

\bigskip

\textbf{ Acknowledgments.} This work was supported in part by grants GO 9746, 10508, 10514, and 10800 from the Space Telescope Science Institute which is operated by AURA under contract from NASA.  Additional support was provide through a NASA Planetary Astronomy grant, NNG04GN31G.  

\bigskip

\centerline\textbf{ REFERENCES}
\bigskip
\parskip=0pt
{\small
\baselineskip=11pt

\refs Agnor, C.~B., and Hamilton, D.~P. (2006)  Neptune's capture of its moon Triton in a binary-planet gravitational encounter.  {\it Nature, 441}, 192-194.

\refs Aitken, R.~G. (1964)  {\it The binary stars}, Dover Publications.

\refs Allen, R. ~L., Gladman, B., Petit, J-M.,  Rousselot, P., Mousis, O.,  Kavelaars, J. ~J., Parker, J., Nicholson, P., Holman, M.,  Doressoundiram, A.,  Veillet, C., Scholl,  H., and Mars, G. (2006) The CFEPS Kuiper Belt Survey: Strategy and Pre-survey Results. {astro-ph: 0510826 v1}.

\refs Astakhov S.~A., Lee E.~A., and Farrelly D. (2005) Formation of Kuiper-belt binaries through multiple chaotic scattering encounters with low-mass intruders.  {\it MNRAS, 360}, 401-415.

\refs Barkume, K.~M., Brown, M.~E., \& Schaller, E.~L. (2006) Water ice on the satellite of Kuiper Belt object 2003 EL$_{61}$.  {\it Astrophys.~J., 640}, L87-L89. 

\refs Bernstein, G.~M., Trilling, D.~E., Allen, R.~L., Brown, M.~E., Holman, M., and Malhotra, R.  (2004) {\it Astrophys.~J. 128}, 1364-1390.

\refs Binzel R.~P., and Hubbard W.~B. (1997)  Mutual events and stellar occultations.  {\it Pluto and Charon}, (Stern, S.~A. and Tholen, D.~J. eds.) pp. 85-102.  University of Arizona Press, Tucson.

\refs Brouwer, D., and Clemence, G.~M. (1961) {\it Methods of Celestial Mechanics}, Academic Press.

\refs Brown M.~E. (2005a) S/2005 (2003 UB\_~313) 1. {\it IAU Circ., 8610}, 1.

\refs Brown M.~E. (2005b) S/2005 (2003 EL\_~61) 2. {\it IAU Circ., 8636}, 1. 

\refs Brown M.~E. (2006)  The largest Kuiper belt objects. {\it Bull.~Amer.~Astron.~Soc., 38}, \#37.01

\refs Brown M.~E., and Calvin W.~M. (2000) Evidence for crystalline water and ammonia ices on Pluto's satellite Charon. {\it Science, 287}, 107-109.

\refs Brown, M.~E. and Trujillo C.~A. (2002) (26308) 1998 SM\_~165. {\it IAU Circ., 7807}, 1. 

\refs Brown, M.~E. and Trujillo C.~A. (2004)  Direct measurement of the size of the large Kuiper belt object (50000) Quaoar.  {\it Astron.~J., 127}, 2413-2417.

\refs Brown, M.~E. and Suer (2007) Satellites of 2003 AZ\_84, (50000), (55637), and (90482). {\it IAU Circ., 8812}, 1.

\refs Brown M.~E., Bouchez A.~H., Rabinowitz D., Sari R., Trujillo C.~A., et al.~(2005a) Keck observatory laser guide star adaptive optics discovery and characterization of a satellite to the large Kuiper Belt object 2003 EL$_{61}$.  {\it Astrophys.~J., 632}, L45-48.

\refs  Brown M.~E., Trujillo C.~A., and Rabinowitz, D. (2005b) 2003 EL\_~61, 2003 UB\_~313, and 2005 FY\_~9. {\it IAU Circ., 8577}, 1. 

\refs Brown M.~E.,  van Dam M.~A., Bouchez A.~H., Le Mignant D., Campbell R.~D., et al.~(2006a)  Satellites of the largest Kuiper Belt objects.  {\it Astrophys.~J., 639}, L43-L46. 

\refs Brown M.~E., Schaller E.~L., Roe H.~G., Rabinowitz D.~L., and Trujillo C.~A. (2006b)  Direct Measurement of the Size of 2003 UB313 from the Hubble Space Telescope.  {\it Astrophys.~J., 643}, L61-64. 

\refs Buie M.~W., Cruikshank D.~P., Lebofsky L.~A. and Tedesco E.~F. (1987) Water frost on Charon. {\it Nature, 329,} 522-523.

\refs Buie, M.~W., Tholen, D.~J., \& Wasserman, L.~H. (1997)  Separate lightcurves of Pluto and Charon. {\it Icarus, 125}, 233-244.

\refs Buie M.~W., and Grundy W.~M. (2000) The distribution and physical state of H2O on Charon. {\it Icarus, 148,} 324-339.

\refs Buie M.~W., Grundy W.~M. , Young E.~F. ,  Young L.~A., and  Stern S.~A.  (2006)  Orbits and photometry of Pluto's satellites: Charon, S/2005 P1, and S/2005 P2.  {\it Astron.\ J., 132}, 290-298.

\refs Canup R.~M. (2005)  A giant impact origin of Pluto-Charon. {\it Science, 307,} 546-550.

\refs Chiang E., Lithwick Y., Murray-Clay R., Buie M., Grundy W., and Holman M.~(2006) A brief history of Trans-Neptunian space.   {\it Protostars and Planets V} (B.~Reipurth et al., eds.), University of Arizona Press, Tucson, in press (astro-ph/0601654).

\refs Christy J.~W., and Harrington R.~S. (1978)  The satellite of Pluto.  {\it Astron.~J., 83,}1005-1008.

\refs Christy J.~W., and Harrington R.~S. (1980)  The discovery and orbit of Charon. {\it Icarus, 44,} 38-40.

\refs Cruikshank D.~P., Pilcher C.~B., and Morrison D.  (1976)  Pluto: Evidence for methane frost. {\it Science, 194,} 835-837.

\refs Cruikshank D.~P., Barucci M.~A., Emery J.~P., Fern‡ndez Y.~R., Grundy W.~M., Noll K.~S., and Stansberry J.~A. (2006)  Physical Properties of Trans-Neptunian Objects. {\it Protostars and Planets V} (B.~Reipurth et al., eds.), University of Arizona Press, Tucson, in press.

\refs Danby J.~M.~A. (1992) {\it Fundamentals of Celestial Mechanics, 2$^{nd}$ Edition}, Willmann-Bell, Inc.

\refs Descamps P. 2005.  Orbit of an astrometric binary system. {\it Celestial Mech.\ and Dynamical Astron., 92,} 381-402.

\refs Dobrovolskis, A.~R., Peale, S.~J., Harris, A.~W. (1997) Dynamics of the Pluto-Charon binary.  in {\it Pluto and Charon} eds. A.~Stern and D.~Tholen, University of Arizona Press, Tucson, pp. 159-190.

\refs Durda, D.~D., Bottke, W.~F., Enke, B.~L., Merline, W.~J., Asphaug, E., Richardson, D., and Leinhardt, Z.~M. (2004)  The Formation of Asteroid Satellites in Large Impacts: Results from Numerical Simulations.  {\it Icarus, 167}, 382-396.

\refs Elliot J.~L.,  Osip D.~J. (2001) 2001 QT\_~297. {\it IAU Circ., 7765}, 4. 

\refs Elliot J.~L., Kern S.~D., Osip D.~J., and Burles S.~M. (2001a) 2001 QT\_~297. {\it IAU Circ., 7733}, 2. 

\refs Elliot J.~L., Kern S.~D., and Clancy K.~B. (2003) 2003 QY\_~90. {\it IAU Circ., 8235}, 2. 

\refs Elliot J.~L., Kern S.~D., Clancy K.~B., Gulbis A.~A.~S., Millis R.~L., Buie M.~W., Wasserman L.~H., Chiang E.~I., Jordan A.~B., Trilling D.~E., and Meech K.~J.  (2005)  The Deep Ecliptic Survey: A search for Kuiper Belt objects and Centaurs. II. Dynamical classification, the Kuiper Belt plane, and the core population. {\it Astrophys.~J., 129}, 1117-1162.

\refs Farinella, P., Milani, A., Nobili, A.~M., Valsecchi, G.~B.  (1979)  Tidal evolution and the Pluto-Charon system. {\it Moon and the Planets, 20}, 415-421.

\refs Fink U., and DiSanti M.~A. (1988) The separate spectra of Pluto and its satellite Charon. {\it Astron.~J., 95,} 229-236.

\refs Funato Y., Makino J., Hut P., Kokubo E., and Kinoshita D. (2004) The formation of Kuiper-belt binaries through exchange reactions. {\it Nature, 427}, 518-520.

\refs Goldreich P., and Soter S. (1966)  Q in the solar system. {\it Icarus, 5}, 375-389.

\refs Goldreich P., Lithwick Y., and Sari R. (2002)  Formation of Kuiper-belt binaries by dynamical friction and three-body encounters.  {\it Nature, 420}, 643-646 ({\it G02}).

\refs Goldreich P., Lithwick Y., and Sari R. (2004) Planet formation by coagulation: A focus on Uranus and Neptune. {\it Ann.~Rev.~Astron.~Astrophys., 42}, 549-601 ({\it G04}).

\refs Grundy W.~M., Noll K.~S., and Stephens D.~C. (2005)  Diverse albedos of small trans-neptunian objects. {\it Icarus, 176}, 184-191. 

\refs Grundy W.~M., Stansberry J.~A., Noll K.~S, Stephens D.~C., Trilling D.~E., Kern S.~D., Spencer J.~R., Cruikshank D.~P., and Levison H.~F. (2007)  The orbit, mass, size, albedo, and density of (65489) Ceto-Phorcys: A tidally-evolved binary Centaur.  {\it Icarus (submitted)}.

\refs Gulbis A.~A.~S., Elliot J.~L., Person M.~J., Adams E.~R., Babcock B.~A., et al. (2006)  Charon's radius and atmospheric constraints from observations of a stellar occultation. {\it Nature, 439}, 48-51.

\refs Harris A.~W. (1998)  A thermal model for near-earth asteroids. {\it Icarus, 131}, 291-301.

\refs Heintz W.~D. (1978) {\it Double Stars}, D. Reidel Publication, Dordrecht.

\refs Hestroffer D., F. Vachier, and B. Balat (2005).  Orbit determination of binary asteroids. {\it Earth, Moon, \& Planets, 97}, 245-260.

\refs Hestroffer D., and Vachier F. (2006)  Orbit determination of binary TNOs.  Paper presented at {\it Trans Neptunian Objects Dynamical and Physical Properties, Catania, Italy, 2006 July 3-7}.

\refs Hubbard W.~B., Hunten D.~M., Dieters S.~W., Hill K.~M., and Watson R.~D.  (1988) Occultation evidence for an atmosphere on Pluto. {\it Nature, 336}, 452-454.

\refs Jewitt, D., and Luu, J. (1998) Optical-infrared spectral diversity in the Kuiper Belt. {\it Astrophys.~J., 115}, 1667-1670. 

\refs Jewitt D.~C., and Sheppard S.~S. (2002) Physical Properties of Trans-Neptunian Object (20000) Varuna. {\it Astron.Z~J., 123}, 2110-2120.

\refs Kavelaars J.~J., Petit J.-M., Gladman B., and Holman M. (2001) 2001 QW\_~322. {\it IAU Circ., 7749}, 1. 

\refs Kenyon S.~J. and Luu J.~X. (1999) Accretion in the early Kuiper Belt. II. Fragmentation. {\it Astron.~J., 118}, 1101-1119.

\refs Kern S.~D. (2006)   A study of binary Kuiper Belt objects.  {\it Ph.D. Thesis} Massachussetts Institute of Technology, Boston.

\refs Kern S.~D., and Elliot J.~L.  (2005)  2005 EO\_~304. {\it IAU Circ., 8526}, 2. 

\refs Kern S.~D., and Elliot J.~L.  (2006)  The frequency of binary Kuiper Belt objects. {\it Astrophys.~J., 643}, L57-L60. 

\refs  Lee M.~H., and Peale S.~J.  (2006)  On the orbits and masses of the satellites of the Pluto-Charon system. {\it Icarus,  184}, 573-583. 

\refs Levison, H.~F., and Stern, S.~A.  (2001) On the size dependence of the inclination distribution of the main Kuiper Belt.  {\it Icarus, 121}, 1730-1735. 

\refs Levison H.~F., Morbidelli A., Gomes R., and Backman D. (2006) Planet migration in planetesimal disks. {\it Protostars and Planets V} (B.~Reipurth et al., eds.), University of Arizona Press, Tucson, in press.

\refs Marchis F., Hestroffer D.,  Descamps P., Berthier J., Laver C., and de~Pater I.  (2005)  Mass and density of asteroid 121 Hermione from an analysis of its companion orbit. {\it Icarus, 178,} 450-464.

\refs Margot, J. ~L., Brown, M. ~E., (2003) A Low-Density M-type Asteroid in the Main Belt. {\it Science, 300}, 1939-1942.

\refs Margot J.~L. (2004)  Binary Minor Planets. {\it Urey Prize Lecture presented at DPS/AAS meeting 2004 November 9, Louisville, KY}.

\refs Margot, J. ~L., Brown, M. ~E., Trujillo, C. ~A., and Sari, R. (2004) HST observations of Kuiper Belt binaries {\it  Bulletin of the American Astronomical Society, 36}, 1081.

\refs Margot J.~L., Brown, M. ~E., Trujillo, C. ~A., Sari, R., Stansberry, J. ~A. (2005)  Kuiper Belt Binaries: Masses, Colors, and a Density {\it Bulletin of the American Astronomical Society, 37}, 737.

\refs McKinnon, W. B. (1984) On the Origin of Triton and Pluto.  {\it Nature, 311}, 355-358.

\refs McKinnon, W. B. (1989) On the Origin of the Pluto-Charon Binary.  {\it Astrophys.~J., 344}, L41-L44.

\refs Merline W.~J., Weidenschilling S.~J., Durda D.~D., Margot J.~L., Pravec P., and Storrs 
A.~D. (2002) Asteroids {\it do} have satellites.  {\it Asteroids III} (Bottke W~F.~Jr., Cellino A., Paolicchi P., and Binzel R.~P. eds.) pp.~289-312.  University of Arizona Press, Tucson. 

\refs Millis R.~L., and Clancy K.~B.  (2003) 2003 UN\_~284. {\it IAU Circ, 8251}, 2. 

\refs Millis R.~L., Wasserman L.~H. ,  Franz O.~G.,  Nye R.~A., Elliot J.~L. , et al.  (1993)  Pluto's radius and atmosphere: Results from the entire 9 June 1988 occultation data set.  {\it Icarus, 105}, 282-297. 

\refs Millis R.~L., Buie M.~W., Wasserman L.~H., Elliot J.~L., Kern S.~D., and Wagner R.~M. (2002) The Deep Ecliptic Survey: A search for Kuiper Belt objects and Centaurs. I. Description of methods and initial results. {\it Astron.~J., 123}, 2083-2109.

\refs Muller, G. ~P.,  Reed R.,  Armandroff, T., Boroson  T. ~A., and Jacoby G. ~H. (1998) What is better than an 8192 x 8192 CCD mosaic imager: Two wide field imagers, one for KPNO and one for CTIO. {\it SPIE Proceedings 3355}, 577-585.

\refs Murray, C.~D., and Dermott, S.~F. (1999) {\it Solar System Dynamics.}  Cambridge University Press, Cambridge.

\refs Nakamura, R., et al. (2000)  Subaru infrared spectroscopy of the Pluto-Charon system. {\it Proc.~Astron.~Soc.~Japan} 52, 551-556.

\refs Noll K.~S. (2003) Transneptunian binaries.  {\it Earth Moon and Planets, 92}, 395-407. 

\refs Noll K.~S. (2006) Solar System binaries. {\it Asteroid, Comets, Meteors 2005, Proceedings IAU Symposium No.~229} (Lazzaro, D., Ferraz-Mello, S., and Fern\'andez, A.~F. eds.) pp.~301-318. Cambridge University Press, Cambridge.

\refs Noll K., Stephens D., Grundy W., Spencer J., Millis R., Buie M., Cruikshank D., Tegler S., and Romanishin W. (2002a) 1997 CQ\_~29. {\it IAU Circ., 7824}, 2. 

\refs Noll K., Stephens D., Grundy W., Spencer J., Millis R., Buie M., Cruikshank D., Tegler S., and Romanishin W. (2002b) 2000 CF\_~105. {\it IAU Circ., 7857}, 1. 

\refs Noll K., Stephens D., Grundy W., Cruikshank D., Tegler S., and Romanishin W. (2002c) 2001 QC\_~298. {\it IAU Circ., 8034}, 1. 

\refs Noll K.~S., Stephens D.~C., Cruikshank D., Grundy W., Romanishin W., and Tegler S. (2003) 1999 RZ\_~253. {\it IAU Circ., 8143}, 1. 

\refs Noll K.~S., Stephens D.~C., Grundy W.~M., and Griffin I. (2004a)  The orbit, mass, and albedo of (66652) 1999 RZ$_{253}$. {\it Icarus, 172,} 402-407.

\refs Noll K.~S., Stephens  D.~C., Grundy W.~M., Osip D.~J., and Griffin I. (2004b)  The orbit and albedo of Trans-Neptunian binary (58534) 1997 CQ$_{29}$. {\it Astron.~J., 128}, 2547-2552. 

\refs Noll K.~S., Grundy W.~M., Stephens D.~C., and Levison H.~F. (2006a)  (42355) 2002 CR\_~46. {\it IAU Circ., 8689}, 1.

\refs Noll K.~S., Grundy W.~M., Levison H.~F., and Stephens D.~C. (2006b)  (60458) 2000 CM\_~114. {\it IAU Circ., 8689}, 2.

\refs Noll K.~S., Grundy W.~M., Stephens D.~C., and Levison H.~F. (2006c)  2003 QW\_~111. {\it IAU Circ., 8745}, 1. 

\refs Noll K.~S., Stephens D.~C., Grundy W.~M., and Levison H.~F. (2006d)  2000 QL\_~251. {\it IAU Circ., 8746}, 1. 

\refs Noll K.~S., Levison H.~F., Stephens D.~C., and Grundy W.~M. (2006e) (120347) 2004 SB\_~60 {\it IAU Circ., 8751}, 1. 

\refs Noll K.~S., Grundy W.~M., Levison H.~F., and Stephens D.~C. (2006f) 1999 RT\_~214. {\it IAU Circ., 8756}, 2.

\refs Noll K.~S., Levison H.~F., Grundy W.~M., and Stephens D.~C. (2006g) Discovery of a binary Centaur. {\it Icarus, 184}, 611-618. 

\refs  Noll K.~S., Grundy W.~M., Levison H.~F., and Stephens D.~C. (2006h) The relative sizes of Kuiper Belt binaries. {\it Bull.~Amer.~Astron.~Soc., 38}, \#34.03.

\refs Noll K.~S., Stephens D.~C., Grundy W.~M., Levison H.~F., and Kern, S.~D. (2007a)  (123509) 2000 WK\_~183. {\it IAU Circ., 8811}, 1. 

\refs Noll K.~S., Kern, S.~D., Grundy W.~M., Levison H.~F., and Stephens D.~C. (2007b)  (119979) 2002 WC\_~19. {\it IAU Circ., 8814}, 1. 

\refs Noll K.~S., Kern, S.~D., Grundy W.~M., Levison H.~F., and Stephens D.~C. (2007c)  2002 GZ\_~31. {\it IAU Circ., 8815}, 1. 

\refs Noll K.~S., Kern, S.~D., Grundy W.~M., Levison H.~F., and Stephens D.~C. (2007d)  2004 PB\_~108 and (60621) 2000 FE\_~8. {\it IAU Circ., 8816}, 1. 

\refs Ortiz J.~L., Guti{\'e}rrez P.~J.,  Casanova V., and Sota A. (2003) A study of short term rotational variability in TNOs and Centaurs from Sierra Nevada Observatory. {\it Astron. and Astrophys., 407}, 1149-1155. 

\refs Osip, D.~J., Kern S.~D., and Elliot J.~L. Physical characterization of the binary Edgeworth-Kuiper Belt object 2001 QT$_{297}$. (2003) {\it Earth Moon and Planets, 92}, 409-421. 

\refs Osip, D. ~J.,  Phillips, D. ~M.,   Bernstein, R.,  Burley, G., Dressler, A. ,  Elliot, J. ~L., Persson, E., Shectman, S. ~A., and Thompson, I. (2004) First-generation instruments for the Magellan telescopes: characteristics, operation, and performance. {\it Ground-based Instrumentation for Astronomy}, (A. F. M. Moorwood and I. Masanori, Ed.), pp. 49-59. 

\refs Peixinho, N., Boehnhardt, H., Belskaya, I., Doressoundiram, A., Barucci, M.~A., and Delsanti, A. (2004) ESO large program on Centaurs and TNOs.  {\it Icarus, 170}, 153-166.

\refs Petit J.-M. and Mousis O. (2004) KBO binaries: how numerous were they? {\it Icarus, 168}, 409-419.

\refs Press W.~H., Teukolsky S~.A., Vetterling W.~T., and Flannery B.~P. (1992) {\it Numerical Recipes in C}, Cambridge University Press, New York.

\refs Rabinowitz D.~L., Barkume K., Brown M.E., Roe H., Schwartz,M., Tourtellotte S., and Trujillo C. (2006)  Photometric observations constraining the size, shape, and albedo of 2003 EL61, a rapidly rotating, Pluto-sized object in the Kuiper belt. {\it Astrophys.~J., 639}, 1238-1251.

\refs Rabinowitz D.~L., Schaefer, B.~E., and Tourtellotte S. (2007)  The diverse solar phase curves of distant icy bodies. I. Photometric observations of 18 Trans-Neptunian objects, 7 Centaurs, and Nereid. {\it Astron.~J., 133}, 26-43.

\refs Richardson D.~C., and Walsh K.~J. (2006) Binary minor planets.  {\it Annual Review of Earth and Planetary Sciences, 34}, 47-81.

\refs Romanishin, W., Tegler, S. ~C., Rettig, T. ~W., Consolmagno, G., and Botthof, B. (2001) 1998 SM165: A large Kuiper belt object with an irregular shape. {\it Publications of the National Academy of Science,  98}, 11863-11866.

\refs Romanishin W., and Tegler S.~C. (2005) Accurate absolute magnitudes for Kuiper Belt objects and Centaurs. {\it Icarus, 179}, 523-526. 

\refs Schaller E.~L., and Brown M.~E. (2003)  A deep Keck search for binary Kuiper Belt objects. {\it Bull.~Amer.~Astron.~Soc., 35},  993.

\refs Sheppard S.~S., and Jewitt D.~C. (2002) Time-resolved photometry of Kuiper Belt objects: Rotations, shapes, and phase functions.  {\it Astron.~J., 124}, 1757-1775. 

\refs Sheppard S.~S. and Jewitt D. (2004) Extreme Kuiper Belt object 2001 QG298 and the fraction of contact binaries. {\it Astron.~J., 127}, 3023--3033.

\refs Smart W.~M. (1980) {\it Textbook on Spherical Astronomy}, Cambridge University Press, New York.

\refs Smith, J.~C., Christy, J.~W., \& Graham, J.~A. (1978) 1978 P 1. {\it IAU Circ., 3241}, 1. 

\refs Spencer J.~R., Stansberry J.~A., Grundy W.~M., and Noll K.~S. (2006) A low density for binary Kuiper belt object (26308) 1998 SM$_{165}$. {\it Bull.~Amer.~Astron.~Soc., 38}, 564. 

\refs Stansberry J.~A., Grundy W.~M., Margot J.~L., Cruikshank D.~P., Emery J.~P., Rieke G.~H., and Trilling D.~E. 2006.  The albedo, size, and density of binary Kuiper belt object (47171) 1999 TC36.  {\it Astrophys.~J., 643}, 556-566.


\refs Stephens D.~C., and Noll K.~S. (2006)  Detection of six trans-Neptunian binaries with NICMOS: A high fraction of binaries in the Cold Classical disk. {\it Astron.~J., 131}, 1142-1148. 

\refs Stephens D.~C., Noll K.~S., Grundy W.~M., Millis R.~L., Spencer J.~R., Buie M.~W., Tegler S.~C., Romanishin W., Cruikshank D.~P.  (2003) HST photometry of trans-Neptunian objects. {\it Earth Moon and Planets, 92}, 251-260. 

\refs Stephens D.~C., Noll K.~S., and Grundy W. (2004) 2000 CQ\_~114. {\it IAU Circ., 8289}, 1.

\refs Stern S.~A. (2002) Implications regarding the energetics of the collisional formation of Kuiper Belt satellites.  {\it Astron.~J. 124}, 2300-2304.

\refs Stern S.~A., Weaver H.~A., Steffl A.~J., Mutchler M.~J., Merline W.~J., Buie M.~W., Young E.~F., Young L.~A., Spencer J.~R.  (2006)  A giant impact origin for Pluto's small moons and satellite multiplicity in the Kuiper belt. {\it Nature, 439}, 946-948. 

\refs Takahashi, S., and Ip, W.-H. (2004)  A Shape-and-Density Model of the Putative Binary EKBO 2001 QG$_{298}$.  {\it Publ.\ Astron.\ Soc.\ Japan, 56}, 1099-1103.


\refs Trujillo C.~A., and Brown M.~E. (2002) 1999 TC\_~36. {\it IAU Circ.,7787}, 1. 

\refs Veillet C., Doressoundiram A., Shapiro J., Kavelaars J.~J., and Morbidelli A. (2001) S/2000 (1998 WW\_~31) 1. {\it IAU Circ., 7610}, 1. 

\refs Veillet C., Parker, J. ~W., Griffin, I, Marsden, B., Doressoundiram, A., Buie, M., Tholen, D. ~J., Connelley, M., Holman, M. ~J. (2002) The binary Kuiper-belt object 1998 WW\_~31. {\it Nature, 416}, 711-713.

\refs Virtanen J., Muinonen K., and Bowell E. (2001)  Statistical ranging of asteroid orbits. {\it Icarus, 154,} 412-431.

\refs Virtanen J., Tancredi G., Muinonen K., and Bowell E. (2003)  Orbit computation for transneptunian objects. {\it Icarus, 161,} 419-430.

\refs Ward W.~R. and Canup R.~M. (2006) Forced resonant migration of Pluto's outer satellites by Charon. {\it Science, 313}, 1107-1109.

\refs Weaver H.~A., Stern S.~A., Mutchler M.~J., Steffl A.~J., Buie M.~J., et al.  (2005) S/2005 P 1 and S/2005 P 2. {\it IAU Circ., 8625}, 1. 

\refs Weaver H.~A., Stern S.~A., Mutchler M.~J., Steffl A.~J., Buie M. (2006) Discovery of two new satellites of Pluto {\it Nature, 439}, 943-945.

\refs Weidenschilling S.~J. (2002) On the origin of binary transneptunian objects.  {\it Icarus, 160}, 212-215.

\refs Young E.~F., and Binzel R.~P. (1994)  A new determination of radii and limb parameters for Pluto and Charon from mutual event lightcurves.  {\it Icarus, 108}, 219-224. 

\refs Young E.~F.., Galdamez K., Buie M.~W., Binzel R.~P., and Tholen D.~J. (1999)  Mapping the variegated surface of Pluto.  {\it Astron.~J., 117}, 1063-1076.

\end{document}